\documentclass[
preprint2
]{aastex}

\usepackage[fleqn]{amsmath}
\usepackage{graphicx}
\usepackage[FIGTOPCAP]{subfigure}
\usepackage{txfonts}
\usepackage{expdlist}
\usepackage{multirow}
\usepackage{natbib}
\bibpunct{(}{)}{;}{a}{}{,}
\usepackage[pdftex=true,colorlinks=true]{hyperref}
\usepackage[all]{hypcap}
\hypersetup{
pdfpagemode = {UseOutlines},
baseurl = {http://www.mpia.de/homes/kuiper},
pdftitle = {On the effects of optically thick gas (disks) around massive stars},
pdfauthor = {Rolf Kuiper},
pdfkeywords = {} 
}

\newcommand{\vONE}{}

\title{On the effects of optically thick gas (disks) around massive stars}

\author{Rolf Kuiper}
\affil{
Jet Propulsion Laboratory,
California Institute of Technology,
4800 Oak Grove Drive,
Pasadena, CA 91109,
USA
}
\email{Rolf.Kuiper@jpl.nasa.gov}
\and
\author{Harold W.~Yorke}
\affil{
Jet Propulsion Laboratory,
California Institute of Technology,
4800 Oak Grove Drive,
Pasadena, CA 91109,
USA
}
\email{Harold.W.Yorke@jpl.nasa.gov}

\begin{abstract}
Numerical simulations have shown that the often cited radiation pressure barrier to accretion onto massive stars can be circumvented, when the radiation field is highly anisotropic in the presence of a circumstellar accretion disk with high optical depth. 
Here, these studies of the so-called flashlight effect are expanded by including the opacity of the innermost dust-free but potentially optically thick gas regions around forming massive stars.
In addition to frequency-dependent opacities for the dust grains, we use temperature- and density-dependent Planck- and Rosseland mean opacities for the gas.
The simulations show that the innermost dust-free parts of the accretion disks are optically thick to the stellar radiation over a substantial fraction of the solid angle above and below the disk's midplane.
The temperature in the shielded disk region decreases faster with radius than in a comparison simulation with 
a lower constant gas opacity, 
and the dust sublimation front is shifted to smaller radii.
The shielding by the dust-free gas in the inner disk thus contributes to an enhanced flashlight effect, which ultimately results in
a smaller opening angle of the radiation pressure driven outflow and in a much longer timescale of sustained feeding of the circumstellar disk by the molecular cloud core.
We conclude that it is necessary to properly account for the opacity of the inner dust-free disk regions around forming 
massive stars in order to correctly assess
the effectiveness of the flashlight effect, 
the opening angle of radiation pressure driven outflows, and
the lifetime and morphological evolution of the accretion disk.
\end{abstract}

\keywords{
ISM: dust, extinction ---
ISM: molecules ---
methods: numerical ---
stars: circumstellar matter ---
stars: formation ---
stars: winds, outflows
\\
\copyright 2012. All rights reserved
}

\begin{document}

\maketitle

\section{Introduction}
\label{sect:introduction}
During their lifetime massive stars can exert a radiative force on their surroundings, which can be 
higher than the gravitational attraction.
How these massive stars can sustain accretion of high-opacity dusty gas was an open question for decades.
Semi-analytical 
\citep{Kahn:1974p799}
and first radiation-hydrodynamical computations in spherical symmetry
\citep{Yorke:1977p376}
supported the idea that the radiative force of a star slows and could ultimately prevent the accretion of material from the molecular core, leading to an upper stellar mass limit of 
$\le 40 \mbox{ M}_\odot$.
\citet{Wolfire:1987p70}
showed how the interstellar medium dust mixture had to be altered to allow the formation of the most massive stars known by spherical symmetric accretion flows.
They found that the absorption probability of the stellar environment has to be reduced to unlikely low values, e.g.~by reducing the dust to gas mass ratio significantly and removing large graphite grains.

Based on the fact that in all of these one-dimensional simulations the collapse was reversed by the re-emitted thermal dust radiation, 
\citet{Nakano:1989p497}
inferred that the radiative force onto the accretion flow can be diminished by an anisotropic optical depth of the environment.
Such an anisotropy can naturally be produced by a circumstellar accretion disk, forming during the pre-stellar core collapse due to angular momentum conservation.

The first numerical study aimed at verifying this idea was by \citet{Yorke:2002p735}, who further expanded on the concept of the so-called ``flashlight effect'' for the anisotropic thermal dust radiation, as introduced by \citet{Yorke:1999p156}.
Whereas \citet{Yorke:1999p156} could verify the significance of the flashlight effect, in particular for estimates of luminosity and spectral appearance, they only considered molecular core masses less than $10 \mbox{ M}_\odot$.
In the \citet{Yorke:2002p735} simulations higher initial core masses up to $120 \mbox{ M}_\odot$ were considered, but the optical depth of the forming accretion disk turned out to be too low to provide a sufficient anisotropy of the radiation field.
As a result, the accretion onto the protostar stopped shortly after the disk formation, limiting the maximum final stellar mass to $42.9 \mbox{ M}_\odot$.
In 
\citet{Krumholz:2009p687}, the authors claimed further feeding of the star and its circumstellar disk by a radiative Rayleigh-Taylor instability in the bipolar cavity regions.
The most massive (and still accreting) protostar formed in this simulation was $41.5 \mbox{ M}_\odot$.

All in all, no star much more massive than the 1D radiation pressure barrier of $M_*^\mathrm{max} \approx 40 \mbox{ M}_\odot$ was formed in these numerical simulations.

In \citet{Kuiper:2012p1151} the radiation transport method used in the \citet{Krumholz:2009p687} study (gray flux-limited diffusion = FLD approximation) was compared with a much more sophisticated hybrid radiation transport approach, namely a combination of a frequency-dependent ray-tracing scheme for the stellar irradiation and a gray FLD solver for the thermal dust (re-)emission.
\citet{Kuiper:2012p1151} showed that the radiation-pressure-dominated cavities became radiative Rayleigh-Taylor unstable only in the runs using the gray FLD approximation for the stellar feedback, whereas they remained stable in the runs using the hybrid approach.
These results were backed up by analytical estimates of the radiative forces acting on the expanding cavity shell.
These results suggest that a radiative Rayleigh-Taylor instability is not a mechanism for overcoming the radiation pressure barrier. 

The next numerical study aimed at verifying the idea of \citet{Nakano:1989p497} regarding the anisotropic radiation field effect was presented by \citet{Kuiper:2010p541}.
The radiation hydrodynamical simulations presented therein confirmed the radiation pressure barrier for spherically symmetric flows and demonstrated that this problem can be circumvented by an anisotropic thermal radiation field, also known as the flashlight effect, as proposed.
It was shown that the flashlight effect even allows the most massive stars known to be formed by disk accretion despite of their strong radiation feedback.  Furthermore, because sufficient angular momentum transport can be provided by gravitational torques forming in the self-gravitating accretion disk \citep{Kuiper:2011p349}, there is no angular momentum barrier to accretion onto the protostar.

Thus, circumstellar accretion disks appear to be a necessary ingredient for massive star formation.
For observational techniques and first disk candidates, we refer the reader to the reviews by
\citet{Zhang:2005p807} on observations 
and
\citet{Cesaroni:2007p8767} on theory and observations of disks around massive stars.

Due to the fact that the optical depth of the dusty regions around massive stars will be highly dominated by dust opacities rather than gas opacities, the numerical studies mentioned above have used simplified gas opacities.
In \citet{Yorke:2002p735}, the dust-free gaseous regions around the forming massive star were -- especially in the disk's midplane -- most likely hidden in the central sink cell of the size of 40, 80, and 160~AU for an initial core mass of 30, 60, and 120~$\mbox{M}_\odot$.
The sink cells were treated as being optically thin.
\citet{Krumholz:2009p687} used a constant Planck- and Rosseland mean gas opacity of $\kappa_\mathrm{P} = \kappa_\mathrm{R} = 0.1 \mbox{ cm}^2\mbox{ g}^{-1}$ and a resolution of the highest refinement level of the AMR grid of 10~AU, i.e.~not resolving the scale height of the dust-free gas disks.
\citet{Kuiper:2010p541, Kuiper:2011p349, Kuiper:2012p1151} used a constant Planck- and Rosseland mean gas opacity of $\kappa_\mathrm{P} = \kappa_\mathrm{R} = 0.01 \mbox{ cm}^2\mbox{ g}^{-1}$ and the resolution of the inner gas disk was about 1~AU at the inner computational boundary, decreasing to larger radii proportional to the radius.

Here, we have improved on this detail by newly implementing the temperature- and density-dependent Planck- and Rosseland mean gas opacities of \citet{Helling:2000p1117}.
For the dust grains we use the frequency-dependent opacities by \citet{Laor:1993p358}, which were used in our previous simulations.
In addition to our default numerical grid of the previous simulations, we here also show the results of extending the computational domain towards a smaller size of the central sink cell of $R_\mathrm{min}=5$~AU, including parts of the dust-free accretion disk around the forming massive star and increasing the polar resolution at the inner boundary to $0.5$~AU.

To our best knowledge, there are no previous dynamical studies of the gaseous disks around massive stars in the literature. 
In \citet{Tanaka:2011p4988}, the authors investigated in analytical and numerical calculations the radiative force by the direct stellar irradiation of the dust sublimation front and in the inner dust-free gas disk for a given static disk configuration.
They propose that the dust sublimation front in the disk's midplane will be shielded by the optically thick inner gas disk.

Here, 
we study this scenario in dynamical simulations of the formation and evolution of circumstellar accretion disks.
In addition to the direct stellar irradiation we include the thermal (re-)emission of both the dusty gas and the dust-free gas and
investigate the dynamical consequences of the optical depth of the dust-free regions for 
the dust sublimation front,
the accretion flow, 
the flashlight effect, and 
the bipolar outflow regions.
In high-mass star forming regions, the radiative force potentially causes or influences the observed wide angle outflows, although it remains unclear, if these outflows are mainly driven by radiation or by a combination of magnetic fields and centrifugal forces.

% TOC:
In Sect.~\ref{sect:methods} we describe the changes made to the numerical code with an emphasis on the newly implemented gas opacities by \citet{Helling:2000p1117}.
Sect.~\ref{sect:sims} provides an overview of the cases considered, their initial conditions, and the numerical configuration.
We present the results of these simulations in Sect.~\ref{sect:results}, giving the physical explanations and discussion of the consequences in Sect.~\ref{sect:discussion}.
We close this report in Sect.~\ref{sect:limitations} with a discussion of the limitations to the current study and an outlook for future investigations.

\section{Methods}
\label{sect:methods}
\begin{figure}[htbp]
\begin{center}
\includegraphics[width=0.48\textwidth]{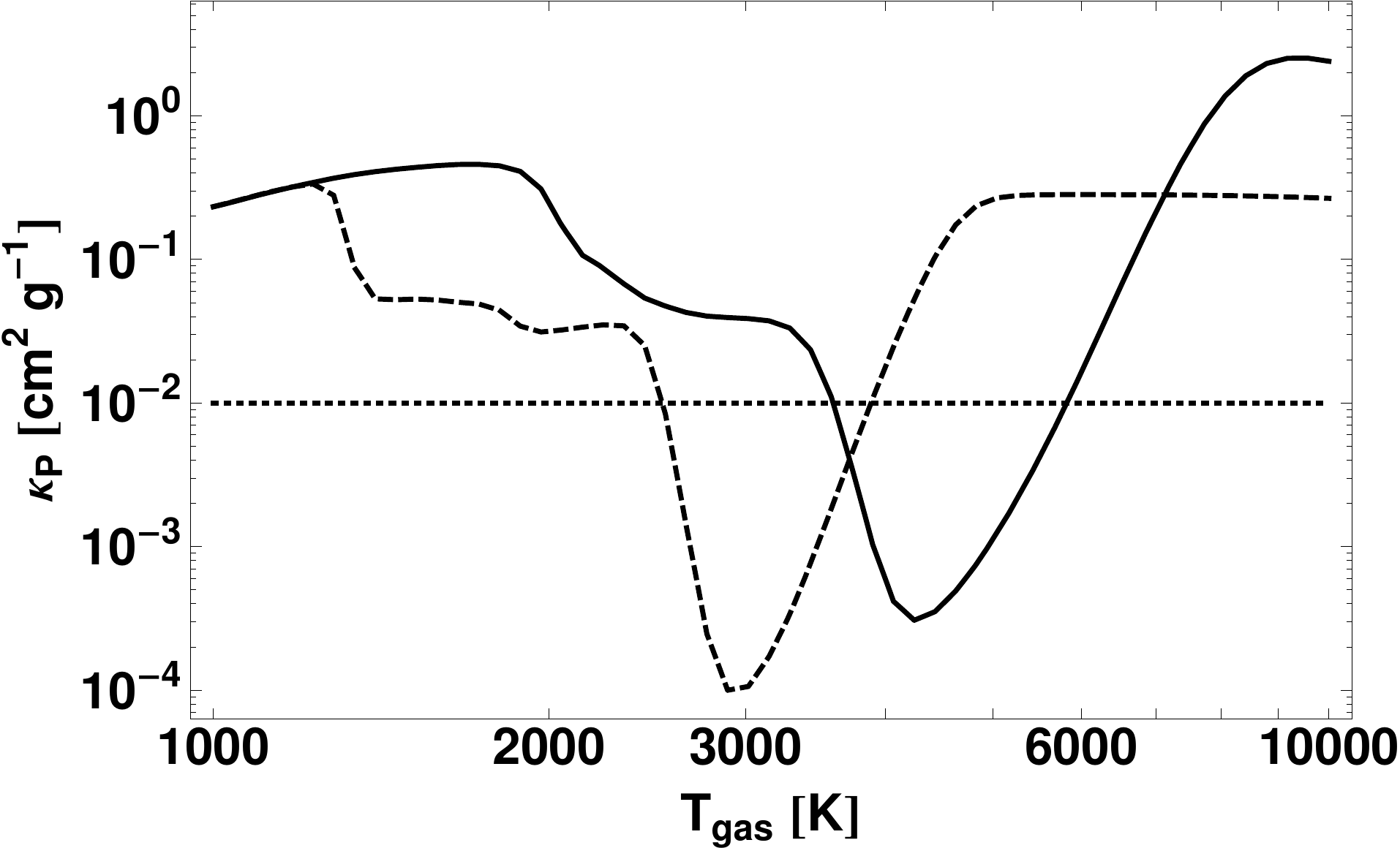}\\
\caption{
Planck mean gas opacities by \citet{Helling:2000p1117} as a function of the gas temperature for two values of the gas density, $\rho = 10^{-11} \mbox{ g cm}^{-3}$ (solid line) and $\rho = 10^{-20} \mbox{ g cm}^{-3}$ (dashed line).
The horizontal line marks the constant gas opacity of $\kappa = 0.01 \mbox{ cm}^2\mbox{ g}^{-1}$ used in the comparison simulation run.
}
\label{fig:kappaP}
\end{center}
\end{figure}
\begin{figure}[htbp]
\begin{center}
\includegraphics[width=0.48\textwidth]{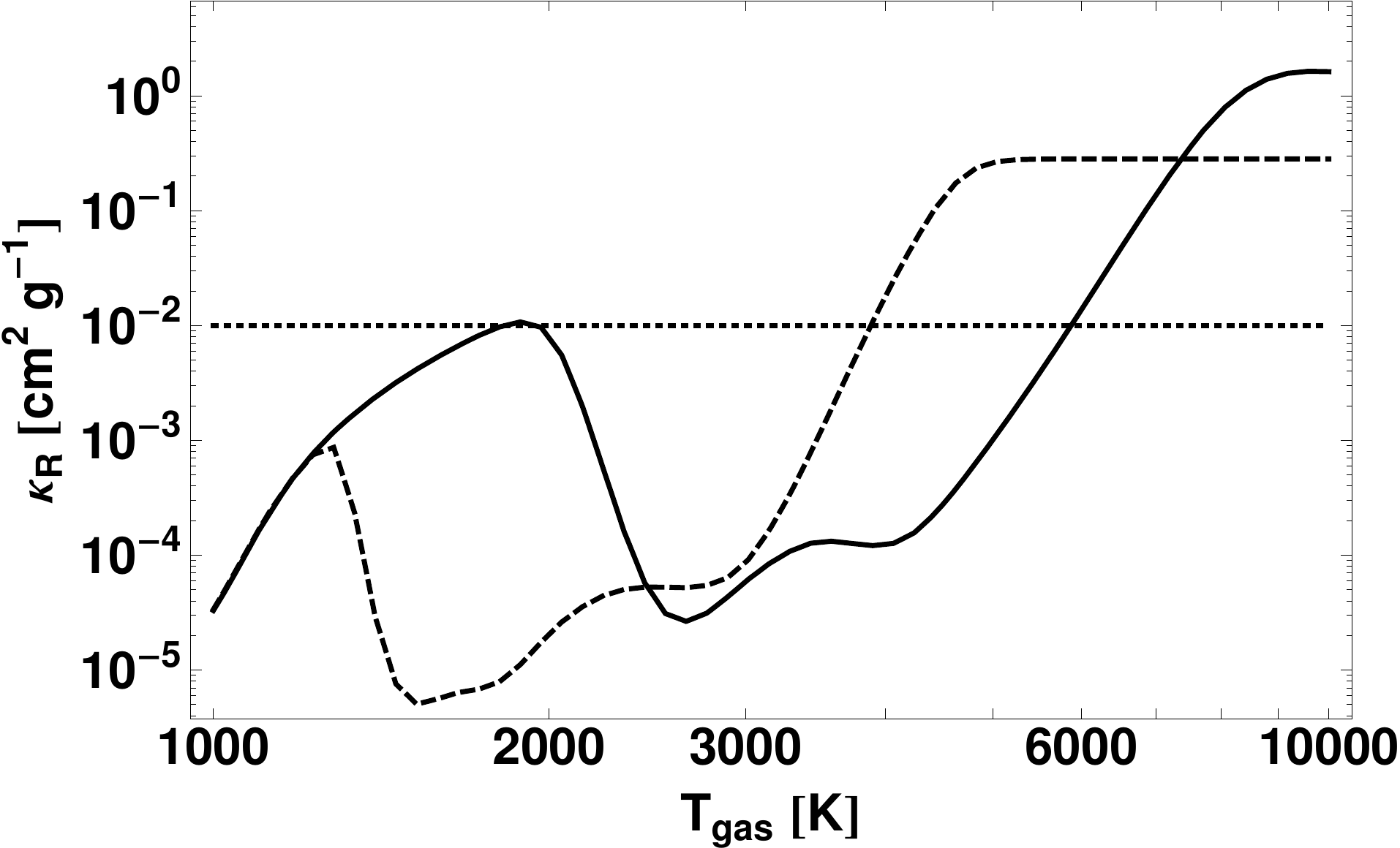}\\
\caption{
Rosseland mean gas opacities by \citet{Helling:2000p1117} as a function of the gas temperature for two values of the gas density, $\rho = 10^{-11} \mbox{ g cm}^{-3}$ (solid line) and $\rho = 10^{-20} \mbox{ g cm}^{-3}$ (dashed line).
The horizontal line marks the constant gas opacity of $\kappa = 0.01 \mbox{ cm}^2\mbox{ g}^{-1}$ used in the comparison simulation run.
}
\label{fig:kappaR}
\end{center}
\end{figure}
The numerical code used to run the pre-stellar core collapse simulations is based on the one used in previous studies \citep{Kuiper:2010p541, Kuiper:2011p349, Kuiper:2012p1151}.
For hydrodynamics we use the open source magneto-hydrodynamics code Pluto \citep{Mignone:2007p544}.
For a detailed description of the derivation, numerical implementation, and benchmarking of the radiation transport solver in use, please see \citet{Kuiper:2010p586}.
The evolution of the central star is computed via fits to the stellar evolutionary tracks of \citet{Hosokawa:2009p23} including the effect of accretion onto forming high-mass stars.
As in our previous studies, the opacity of dust grains is computed taking into account the frequency-dependence for the stellar irradiation and using frequency-averaged (gray) Planck- and Rosseland mean opacities for the thermal dust emission.
We use the frequency-dependent dust opacities by \citet{Laor:1993p358}.

Here, we improved the radiation transport by implementing gray, temperature- and density-dependent Planck- and Rosseland mean opacities of the gas.
Whereas dust grains will dominate the opacity in major parts of the collapsing pre-stellar core, gas opacities could become important in the dust-free regions close to the protostar.
Here, we have implemented the gas opacities computed in \citet{Helling:2000p1117}. 
These gas opacities were also used in a study deriving opacities for a broad parameter range (in density and temperature) for proto-planetary disks in general \citep{Semenov:2003p79}.
Opacity values for given logarithmic gas temperatures and densities are interpolated bi-linearly from a $71 \times 71$ grid.
The Planck- and Rosseland mean gas opacities as a function of gas temperature are depicted in Figs.~\ref{fig:kappaP} and \ref{fig:kappaR} for two values of gas density.
The optical depth of regions of lower temperatures than plotted will most likely be dominated by dust grains.

\section{Simulations overview, initial conditions, and numerical configuration}
\label{sect:sims}
We have performed three different simulations:
For the ``fiducial run'' we computed the collapse of a pre-stellar core including the newly implemented gas opacities.
A ``comparison run'' with the same initial conditions and numerical configuration is performed using a constant gas opacity of $\kappa = 0.01 \mbox{ cm}^2\mbox{ g}^{-1}$.
To determine the dependence on the extent of the computational domain towards the direction of the central star, the fiducial run is repeated (called the ``convergence run'') using a smaller sink cell size of $R_\mathrm{min} = 5$~AU instead of the value $R_\mathrm{min} = 10$~AU used in the fiducial and comparison runs. 
An overview of these simulations is given in Table~\ref{tab:runs}.

The initial condition of the pre-stellar cores is given by 
an outer core radius of $0.1~\mbox{pc}$, 
a mass of $100 \mbox{ M}_\odot$, and 
a temperature of $20~\mbox{K}$.
The core is initially in rigid solid body rotation.

The numerical grid in use is identical to several of our previous studies.
The resolution of the spherical grid at the inner computational boundary is about $1~\mbox{AU}$ in the radial as well as in the polar direction, assuming axial symmetry.
The polar extent of the grid covers the angle from $0\degr$ to $90\degr$, assuming midplane symmetry.
Towards the outer computational boundary, the grid cell size increases linearly with the radius.
In the run using the smaller central sink cell size of $R_\mathrm{min} = 5$~AU, the radial grid is extended towards the smaller radii using a uniform step size of 1~AU.
For more details on the numerical grid, the numerical solvers, and the sub-grid models such as shear viscosity and the dust model, please see \citet{Kuiper:2010p586} and \citet{Kuiper:2010p541}.

\begin{table}[tbhp]
\begin{tabular}{l | c c | c}
Run tag & $R_\mathrm{min}$ [AU] & $\kappa_\mathrm{gas}$ & $M_* \mbox{ [M}_\odot\mbox{]}$ \\
\hline
fiducial run  & 10 & H & \hspace{2mm}$60.5^+$ \\
comparison run  & 10 & c & \hspace{2mm}52.1\hspace{2mm} \\
convergence run  & \hspace{1.1ex}5 & H & \hspace{2mm}$58.0^+$ \\ 
\end{tabular}
\caption{
Overview of simulations performed.
From left to right the columns denote the 
name tag,
the radius of the inner boundary condition, and 
the gas opacities in use:
``H'' refers to the usage of the frequency-averaged Planck- and Rosseland mean gas opacities by \citet{Helling:2000p1117},
``c'' refers to the usage of a constant gas opacity of $0.01 \mbox{ cm}^2 \mbox{ g}^{-1}$.
The last column denotes the resulting final mass of the star forming in the central sink cell.
For the fiducial and convergence run the ``+'' sign denotes the fact that in the end of the simulation the star is still accreting.
The remaining mass in the computational domain in the fiducial run is $4.0 \mbox{ M}_\odot$ at $t=253$~kyr.
The remaining mass in the computational domain in the convergence run is $20.8 \mbox{ M}_\odot$ at $t=80$~kyr.
}
\label{tab:runs}
\end{table}

\section{Results}
\label{sect:results}
\begin{figure}[htbp]
\begin{center}
\includegraphics[width=0.48\textwidth]{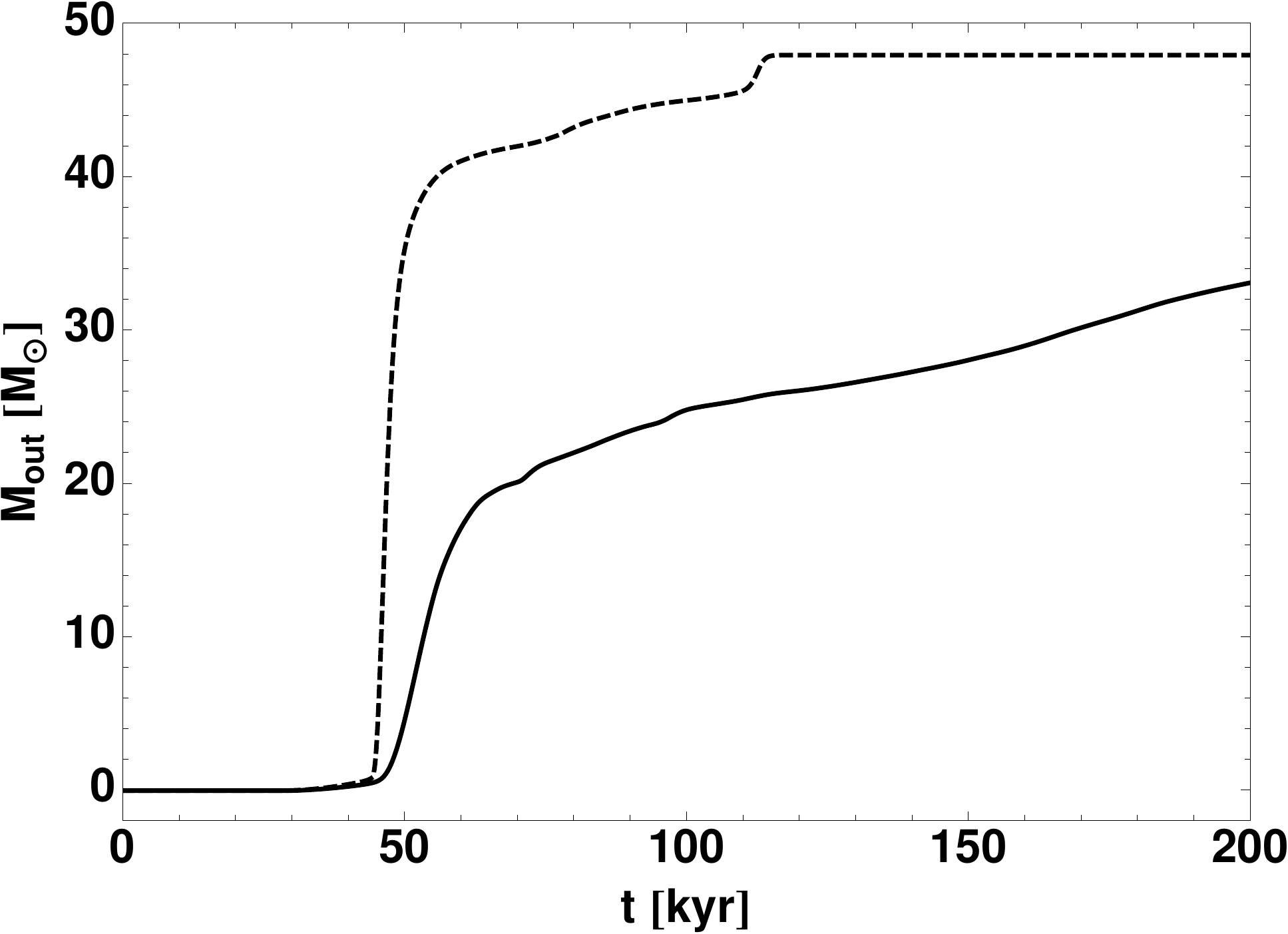}\\
\caption{
Mass loss of the pre-stellar molecular cores due to radiative acceleration as a function of time.
The solid line denotes the fiducial simulation run with gas opacities by \citet{Helling:2000p1117}.
The dashed lines denotes the comparison run using a constant gas opacity.
}
\label{fig:Mout_vs_t}
\end{center}
\end{figure}

\begin{figure}[htbp]
%\begin{center}
\hspace{4.5mm}
\includegraphics[width=0.440\textwidth]{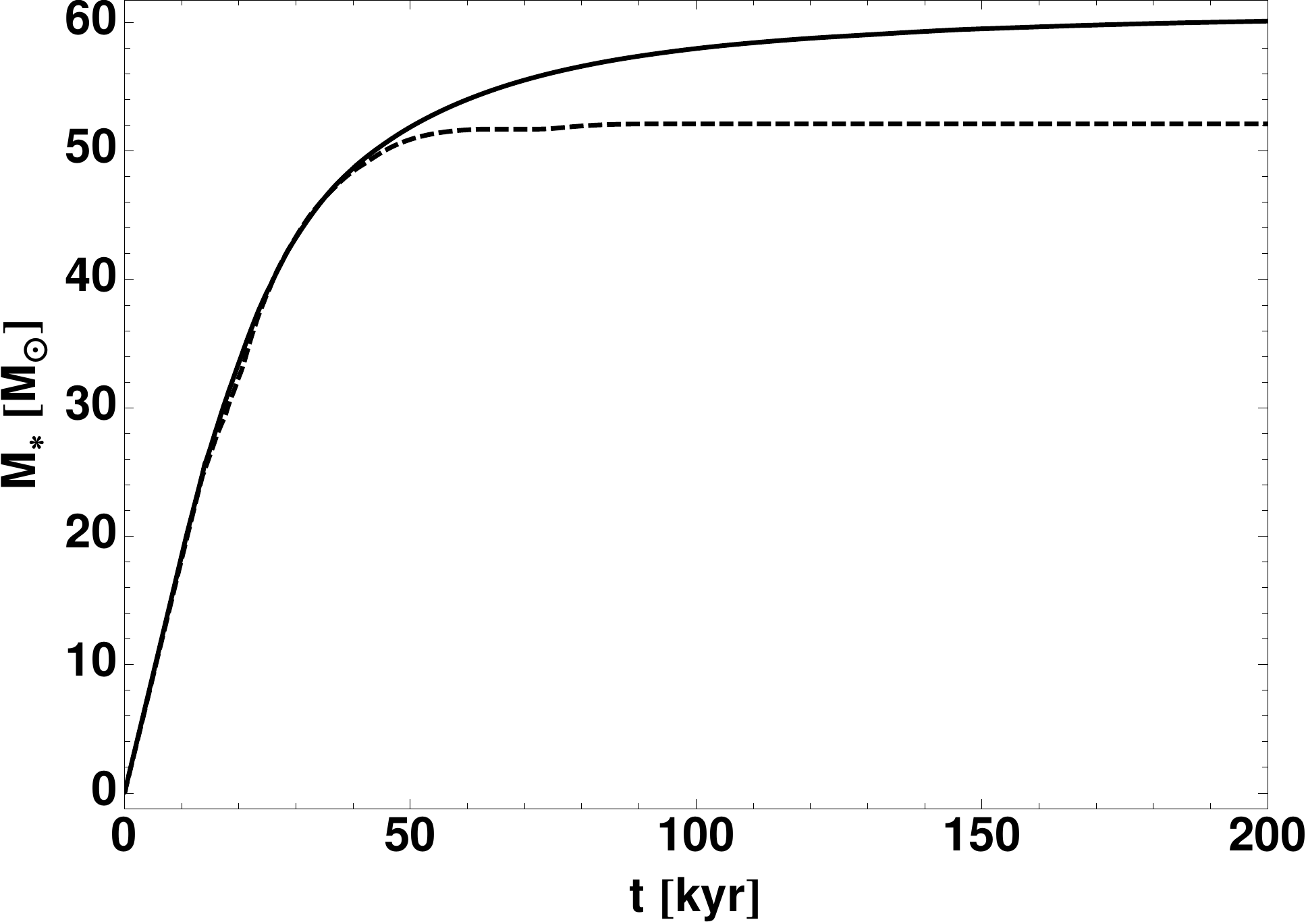}\\
\includegraphics[width=0.48\textwidth]{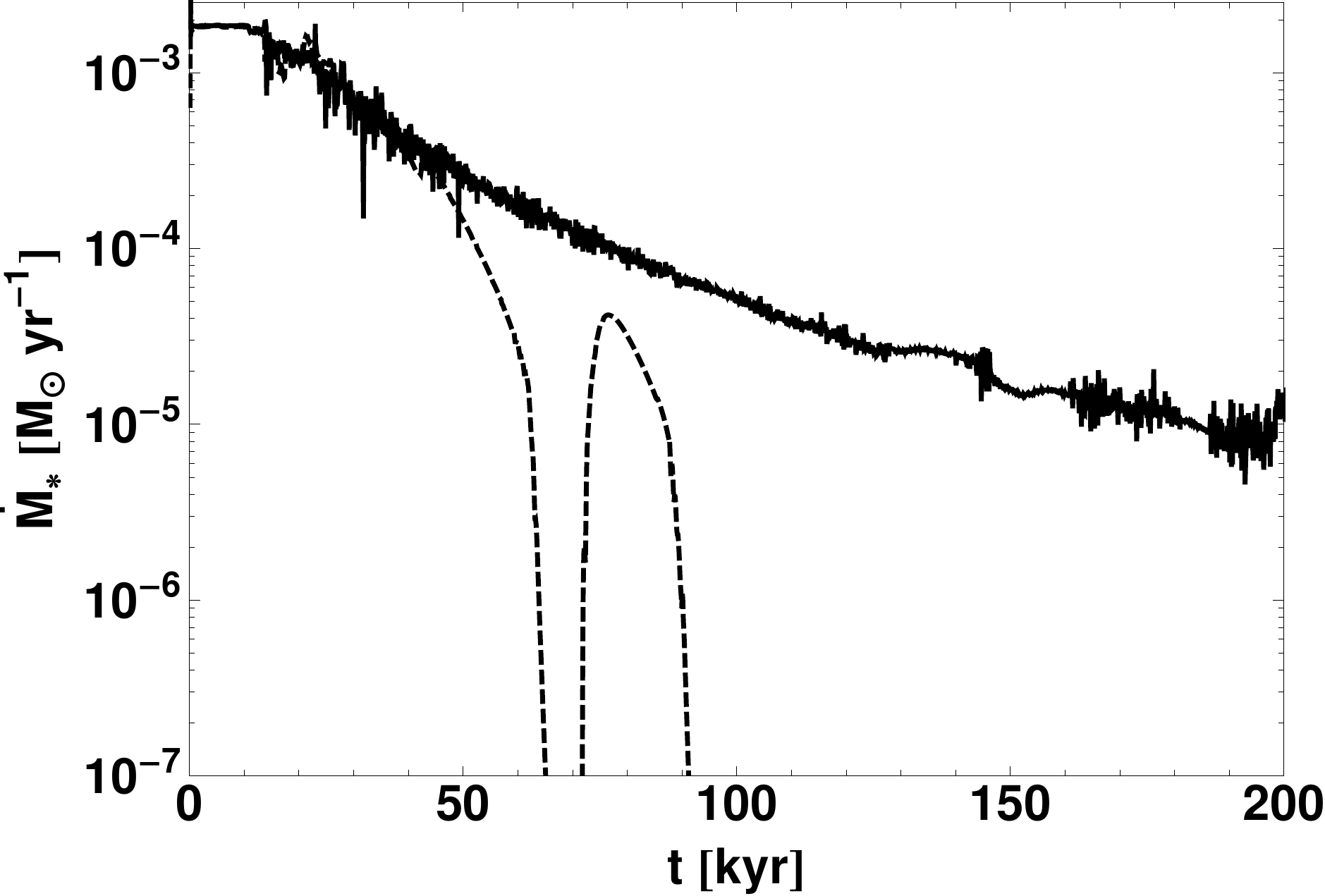}\\
\caption{
Mass of the central star (upper panel) and accretion rate onto the star (lower panel) as a function of time.
The solid lines denote the fiducial simulation run with the gas opacities by \citet{Helling:2000p1117}.
The dashed lines denote the comparison run using a constant gas opacity.
}
\label{fig:Mdot_vs_t}
%\end{center}
\end{figure}

During the initial free-fall epoch and the early epoch of disk formation and evolution, the fiducial, comparison, and convergence runs show no qualitative and only minor quantitative differences.
The central star eventually becomes so luminous that its radiative force exerted onto the environment overcomes gravity.
Over time, these radiative forces lead to significant mass loss from the pre-stellar molecular core.
The evolution of the simulation runs using different gas opacities starts to deviate significantly once the radiation-pressure-dominated cavities are formed.
First, the opening angle of the radiation-driven outflow is smaller for the run with the newly implemented gas opacities, resulting
in different mass loss rates from the pre-stellar core, see Fig.~\ref{fig:Mout_vs_t}.
This mass loss rate is computed by integrating the mass flux through the outer computational boundary at $R_\mathrm{max} = 0.1$~pc.
This boundary is treated as a semi-permeable wall, i.e.~material with outward radial velocity is allowed to leave the computational domain, but inflow of material is prohibited.
The fiducial run with the newly implemented gas opacities initially has a much lower mass loss rate.
Second, the fiducial and the comparison run have very different accretion histories of the forming star, see Fig.~\ref{fig:Mdot_vs_t}.
The fiducial run has a much longer epoch of sustained accretion than in the comparison run (see explanation in Sect.~\ref{sect:discussion}).

{
\vONE
Radiative acceleration stops disk accretion in the comparison run at about 65~kyr.
After this initial shut off the mass in the remnant disk piles up and leads to an epoch of low accretion (a few $10^{-5} \mbox{ M}_\odot \mbox{ yr}^{-1}$) from roughly 70 to 90~kyr.
Eventually, the radiation force is able to disperse the remnant disk and accretion shuts off. 
}

\section{Discussion}
\label{sect:discussion}
In the following, we discuss the reason for the different opening angles of the radiation driven outflow and the associated different mass loss rates out of the pre-stellar molecular cores (Sect.~\ref{sect:core}) and the reasons for a longer epoch of sustained disk accretion when gas opacities are properly taken into account (Sect.~\ref{sect:disk}).

\subsection{Envelope evolution}
Within the low-density optically thin bipolar regions the choice of gas opacities has little dynamical effect.
The shape and extent of the low-density bipolar region, however, depends strongly on the morphology of the accretion disk, 
where -- close to the mid-plane of the dust-free zone -- the proper treatment of gas opacities is indeed of importance.
\label{sect:core}
\begin{figure}[htbp]
\begin{center}
\includegraphics[width=0.48\textwidth]{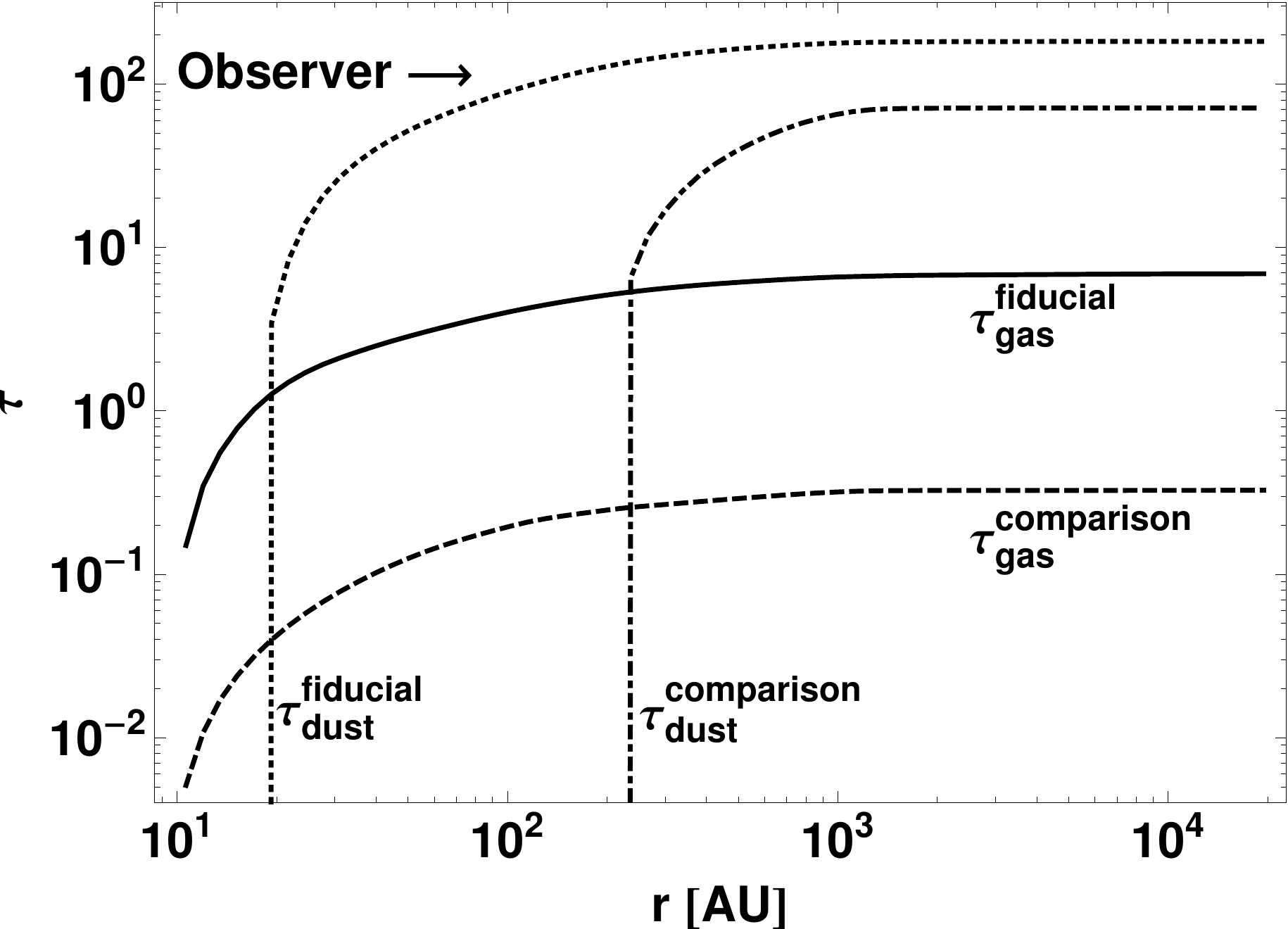}\\
\caption{
Integrated optical depth of the stellar radiation at an angle of $15\degr$ above the disk's midplane
from the central star to the outer boundary at $t=45$~kyr.
{\vONE
The optical depth is computed using Planck mean gas and dust opacities.
}
The solid and dotted lines denote the optical depths of the gas and dust phase, respectively, in the fiducial simulation run using the gas opacities by \citet{Helling:2000p1117}, whereby the optical depth is integrated radially from the central star to the outer computational boundary.
The dashed and dot-dashed lines denote the optical depths of the gas and dust phase, respectively, in the comparison run with a constant gas opacity.
}
\label{fig:tau}
\end{center}
\end{figure}
In the fiducial run at $t=45$~kyr, for example, the dust-free region around the forming massive star becomes optically thick to the stellar radiation at angles of $\theta \approx \pm 18\degr$ above and below the disk's midplane.
At this point in time, the fiducial and the comparison runs differ in their optical depths to the stellar radiation (see Fig.~\ref{fig:tau}), resulting in differences of the corresponding pre-stellar cores' mass loss and their disks' evolution.
The gas in the sink cell itself is assumed to be optically thin; for further information please see the appendix for a comparison to the convergence simulation run using a smaller central sink cell.

\begin{figure*}[htbp]
%\begin{center}
\hspace{10mm}
\vspace{5mm}
\includegraphics[width=0.92\textwidth, angle=270,totalheight=0.1\textheight]{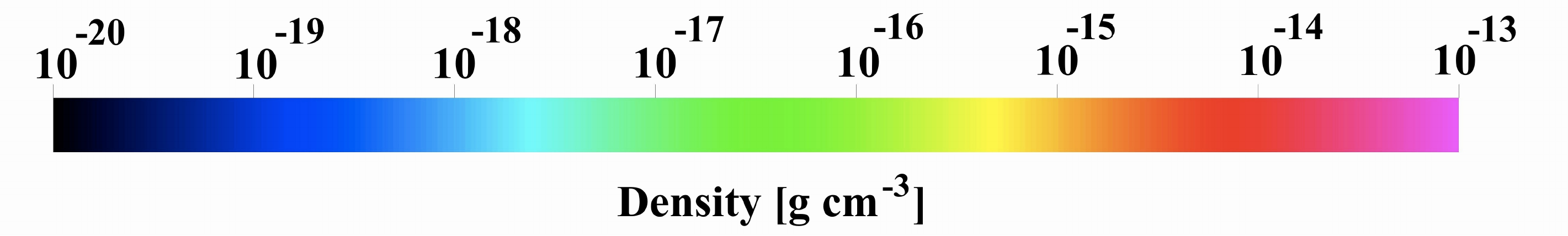}\\
\vspace{5mm}
\includegraphics[width=0.96\textwidth, angle=270,totalheight=0.3\textheight]{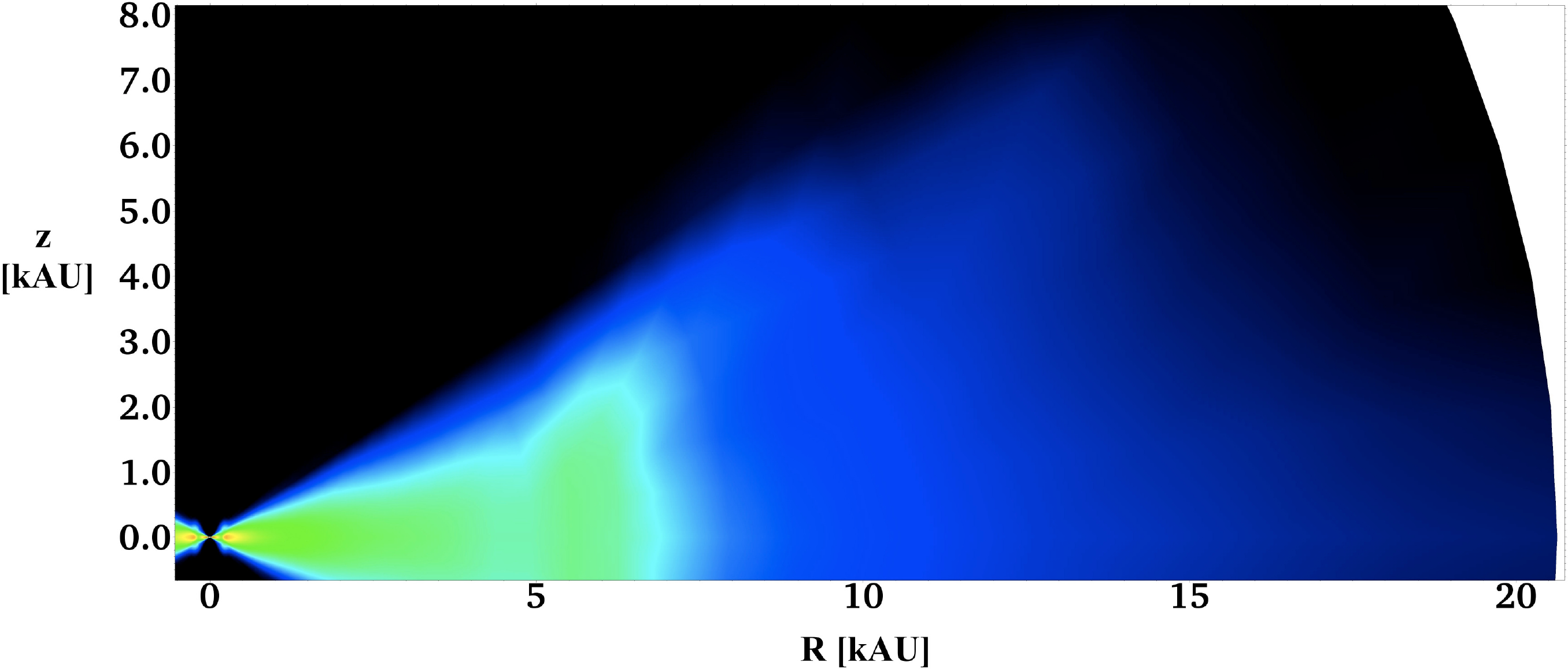}\\
\includegraphics[width=0.96\textwidth, angle=270,totalheight=0.3\textheight]{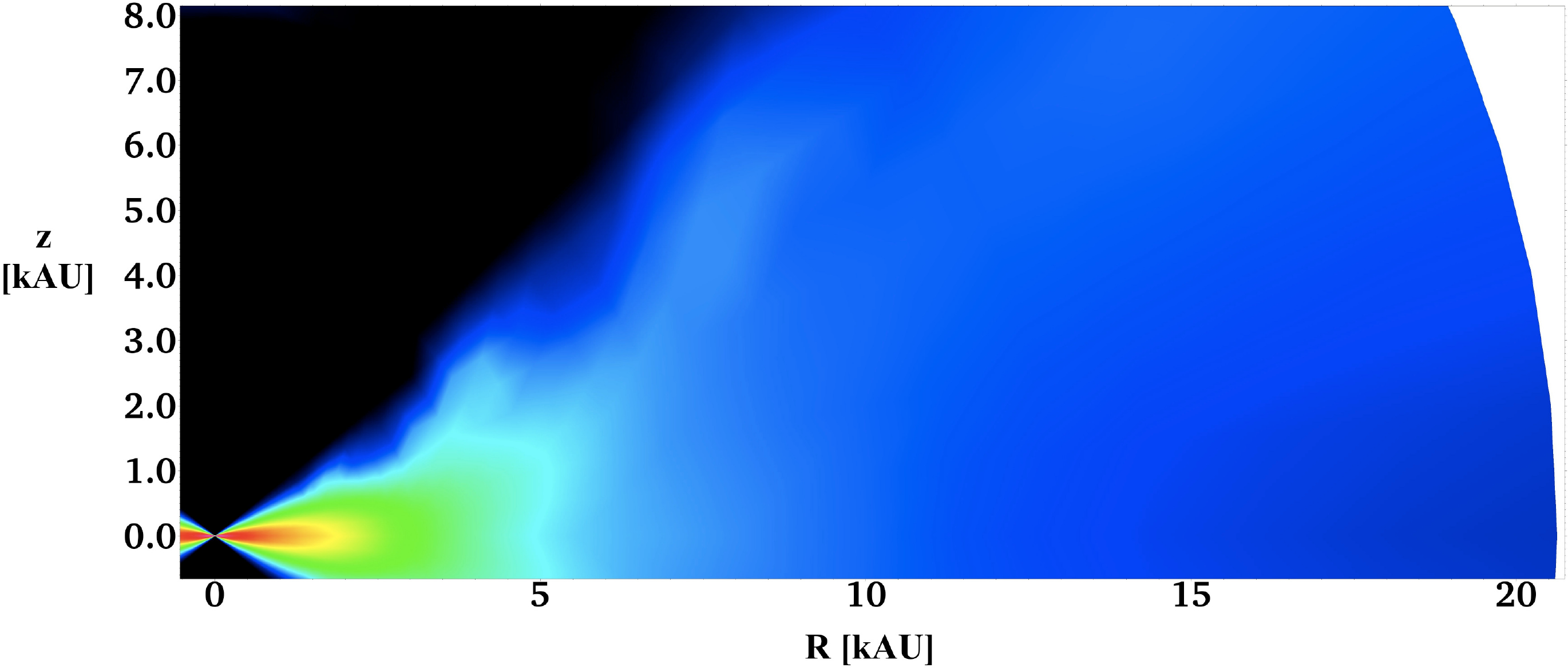}\\
\caption{
Gas density distribution of the disk feeding envelope region at $t=63$~kyr in the comparison run (upper panel) and the fiducial simulation run (lower panel).
The snapshot in time is chosen shortly prior to the dispersion of the accretion disk in the comparison run.
}
\label{fig:visit}
%\end{center}
\end{figure*}

Because the gas is optically thick to stellar radiation $15\degr$ above and below the disk's midplane, 
dust grains persist, where otherwise (without the optically thick gas) they would sublimate due to heating by direct stellar radiation.
The radius of dust sublimation for an angle of $15\degr$ above the disk's midplane at this point in evolution decreases from more than 200~AU in the comparison run to less than 20~AU in the fiducial run.
The net effect is to increase the optical depth of the innermost 200~AU along the line of sight by more than two orders of magnitude.
{\vONE
Although it is difficult to define precisely the "opening angle" of the outflow cavity - it varies both spatially and temporally - we note that the large differences of optical depth in the inner parts of the envelope lead to an apparent}
smaller opening angle of the outflow in the fiducial simulation run (see Fig.~\ref{fig:visit}), which results in a smaller mass loss rate of the pre-stellar core.

{\vONE
The opening angles of the outflow in the fiducial run with proper gas opacities, are more consistent with observations, whereby molecular outflows tend to have opening angles of $90\degr$ within 50 AU of the star
\citep{Arce:2007p10514}.
}

\subsection{Disk evolution}
\label{sect:disk}
\begin{figure*}[p]
\begin{center}
\subfigure[25~kyr]{
\includegraphics[width=0.48\textwidth]{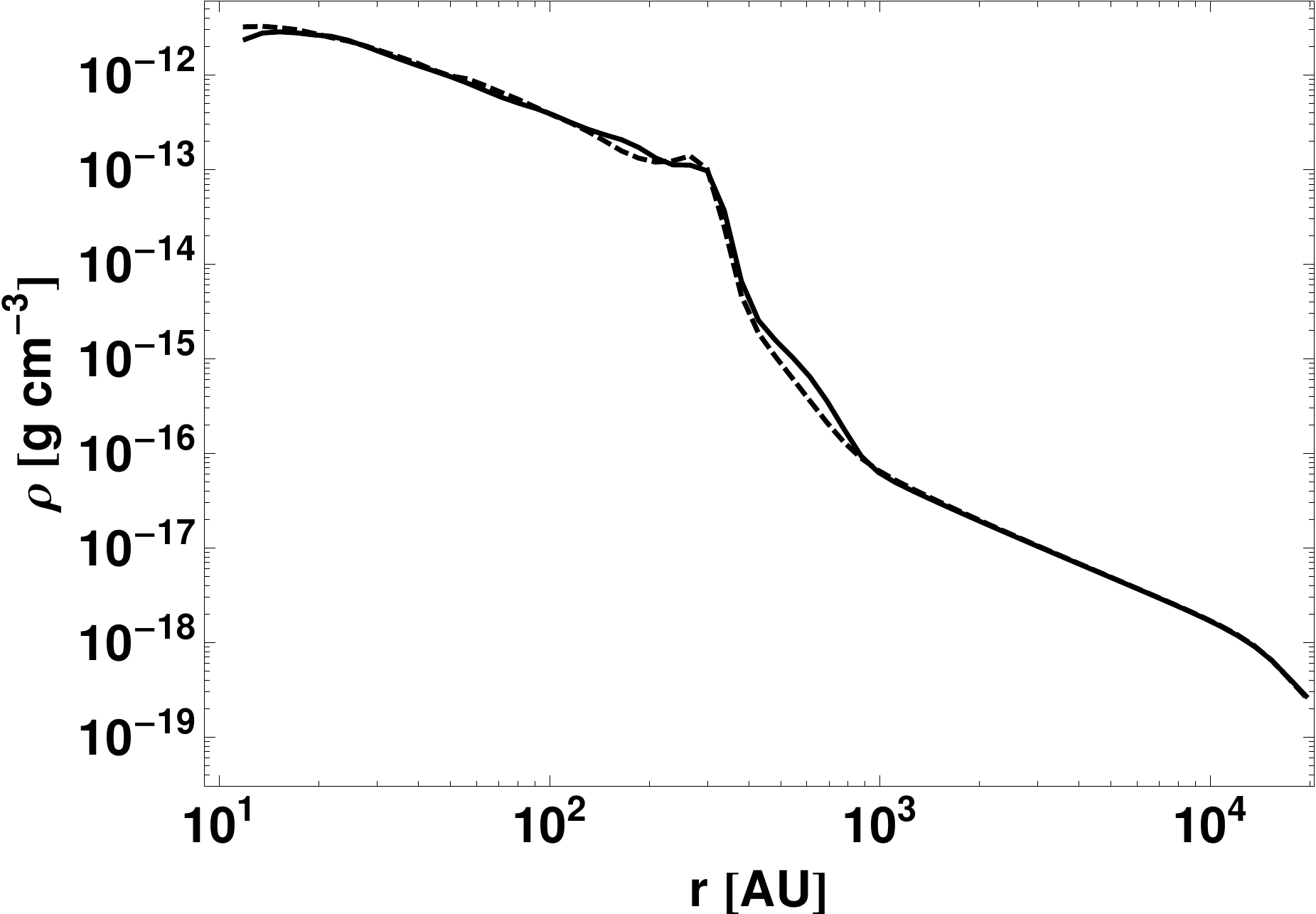}
\label{fig:rho25}
}
\subfigure[40~kyr]{
\includegraphics[width=0.48\textwidth]{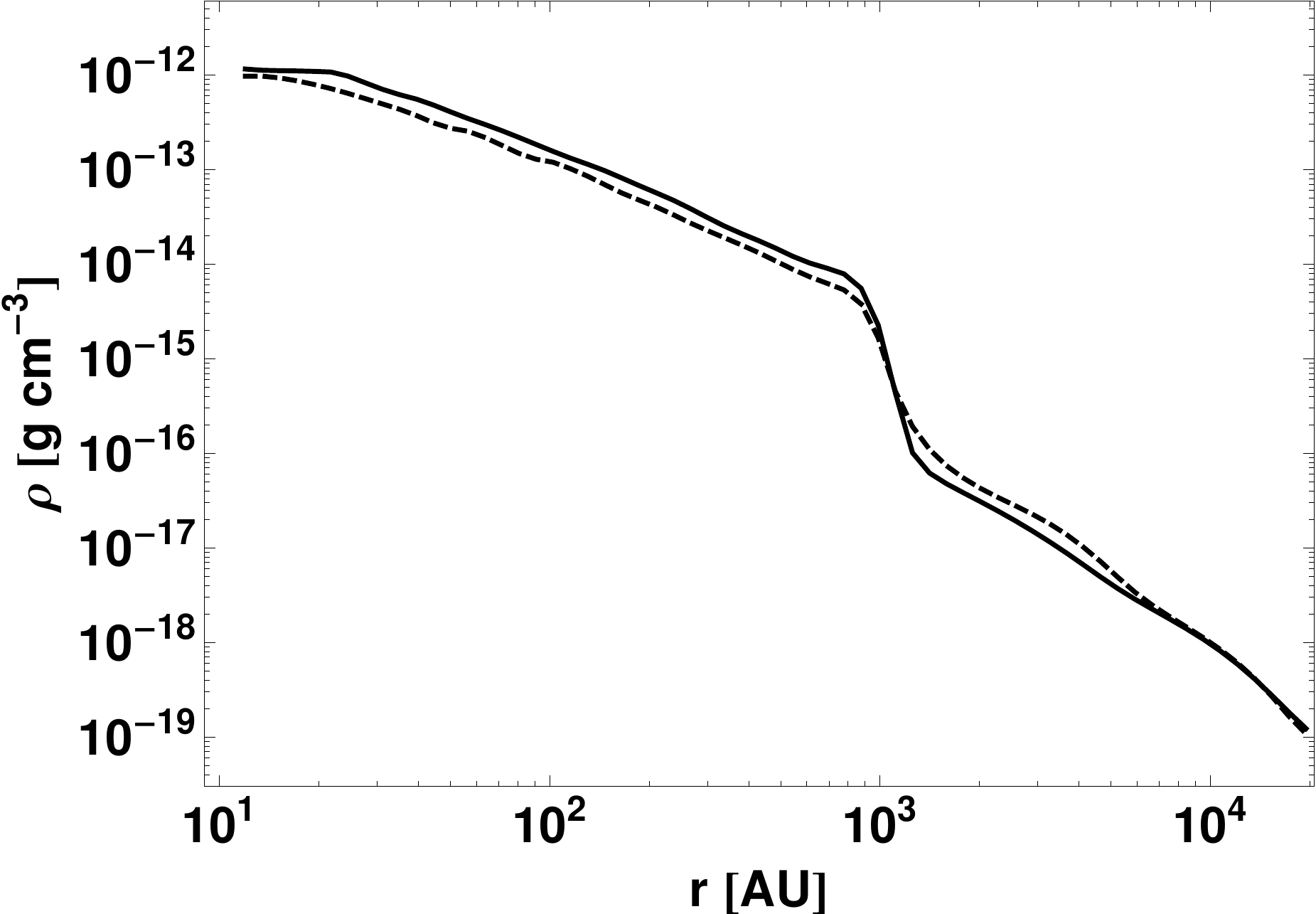}
\label{fig:rho40}
}
\subfigure[50~kyr]{
\includegraphics[width=0.48\textwidth]{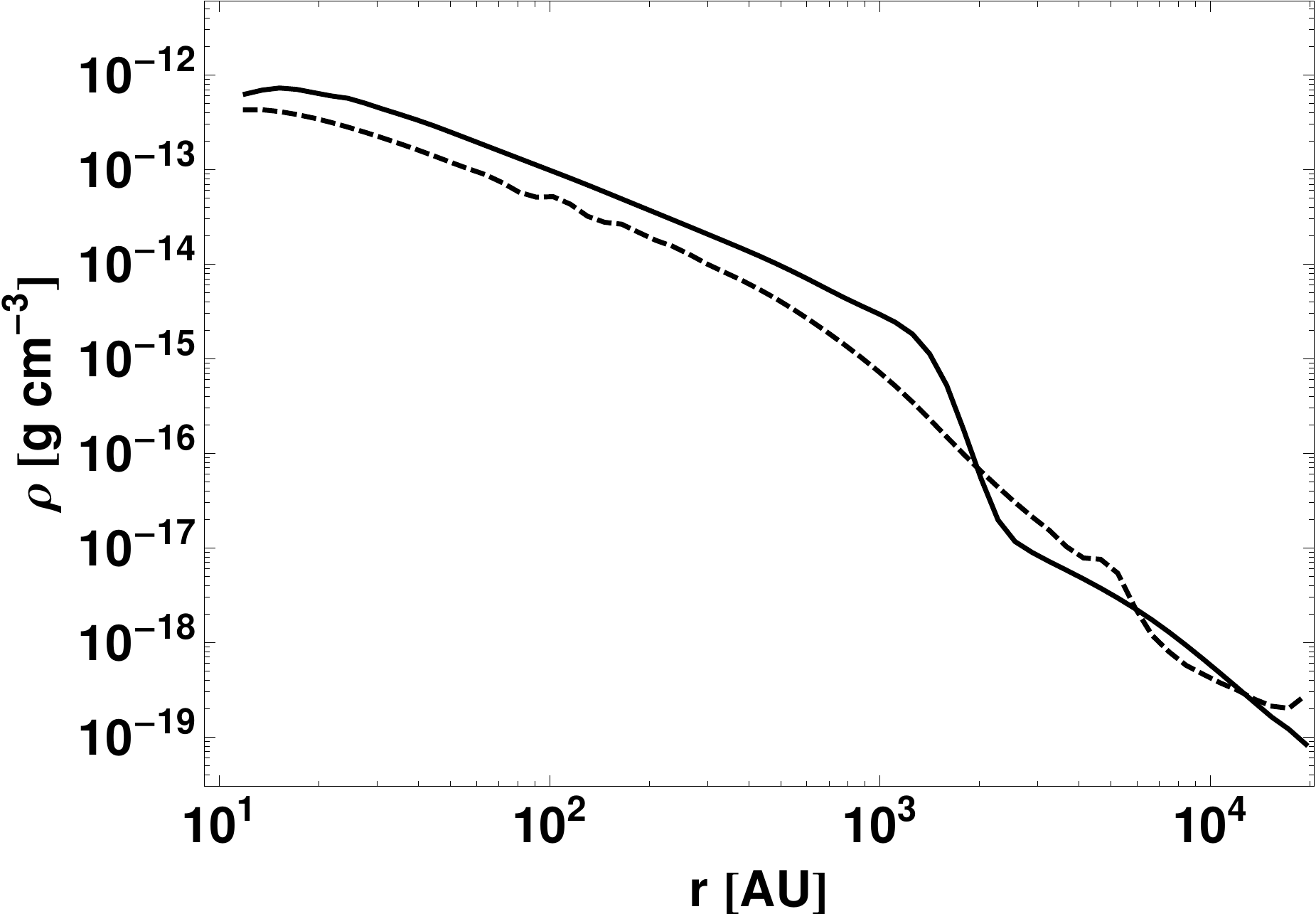}
\label{fig:rho50}
}
\subfigure[55~kyr]{
\includegraphics[width=0.48\textwidth]{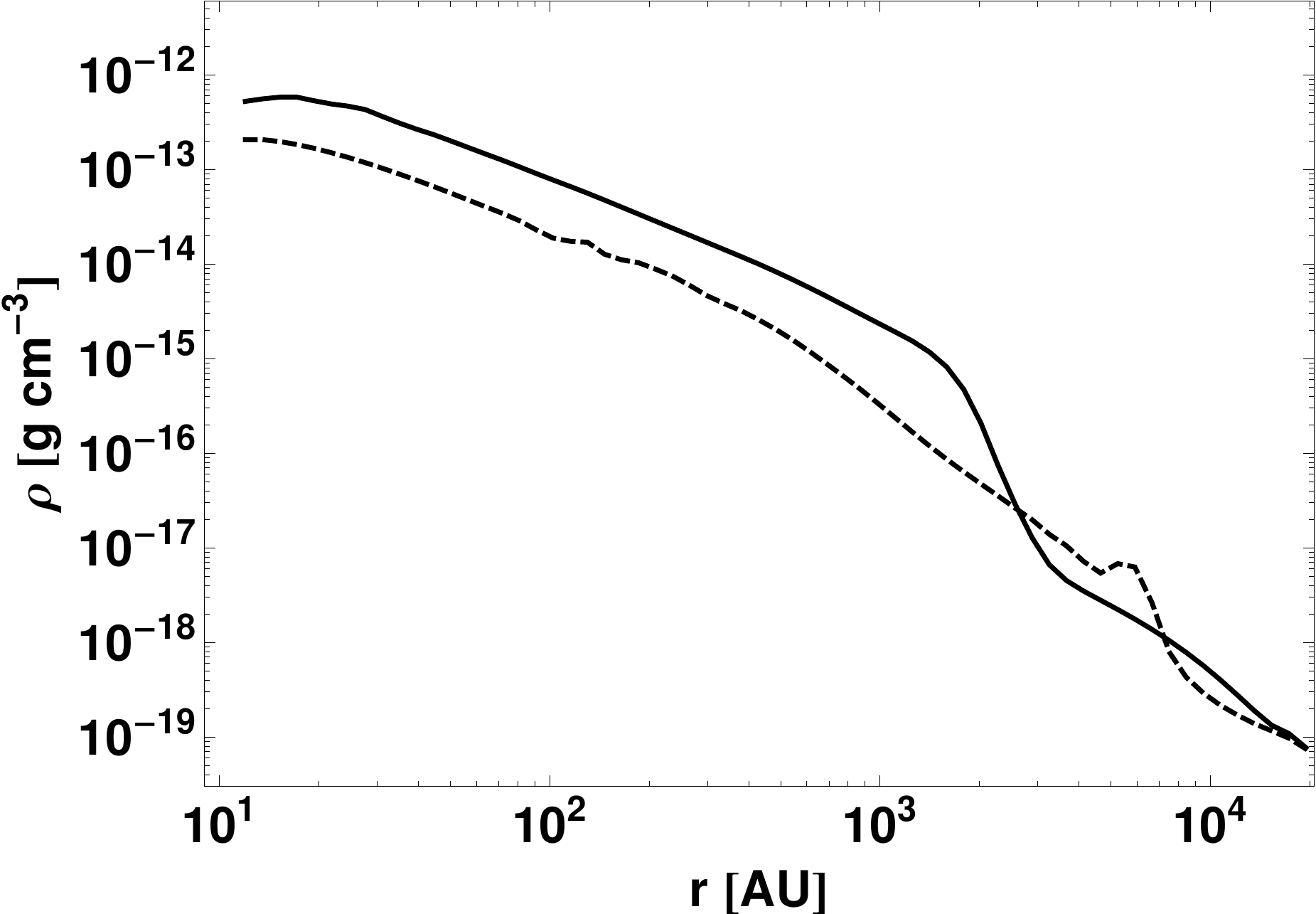}
\label{fig:rho55}
}
\subfigure[60~kyr]{
\includegraphics[width=0.48\textwidth]{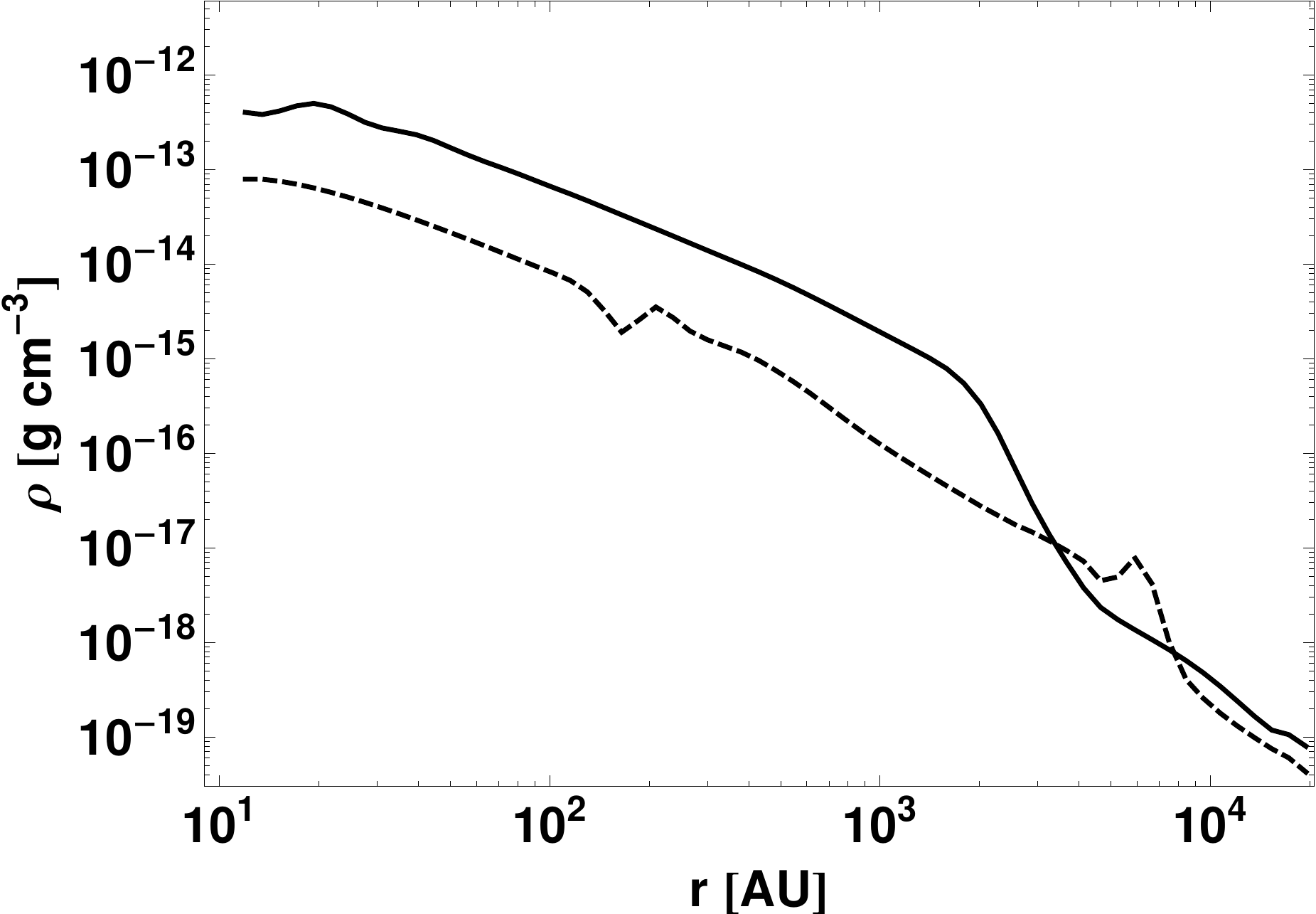}
\label{fig:rho60}
}
\subfigure[65~kyr]{
\includegraphics[width=0.48\textwidth]{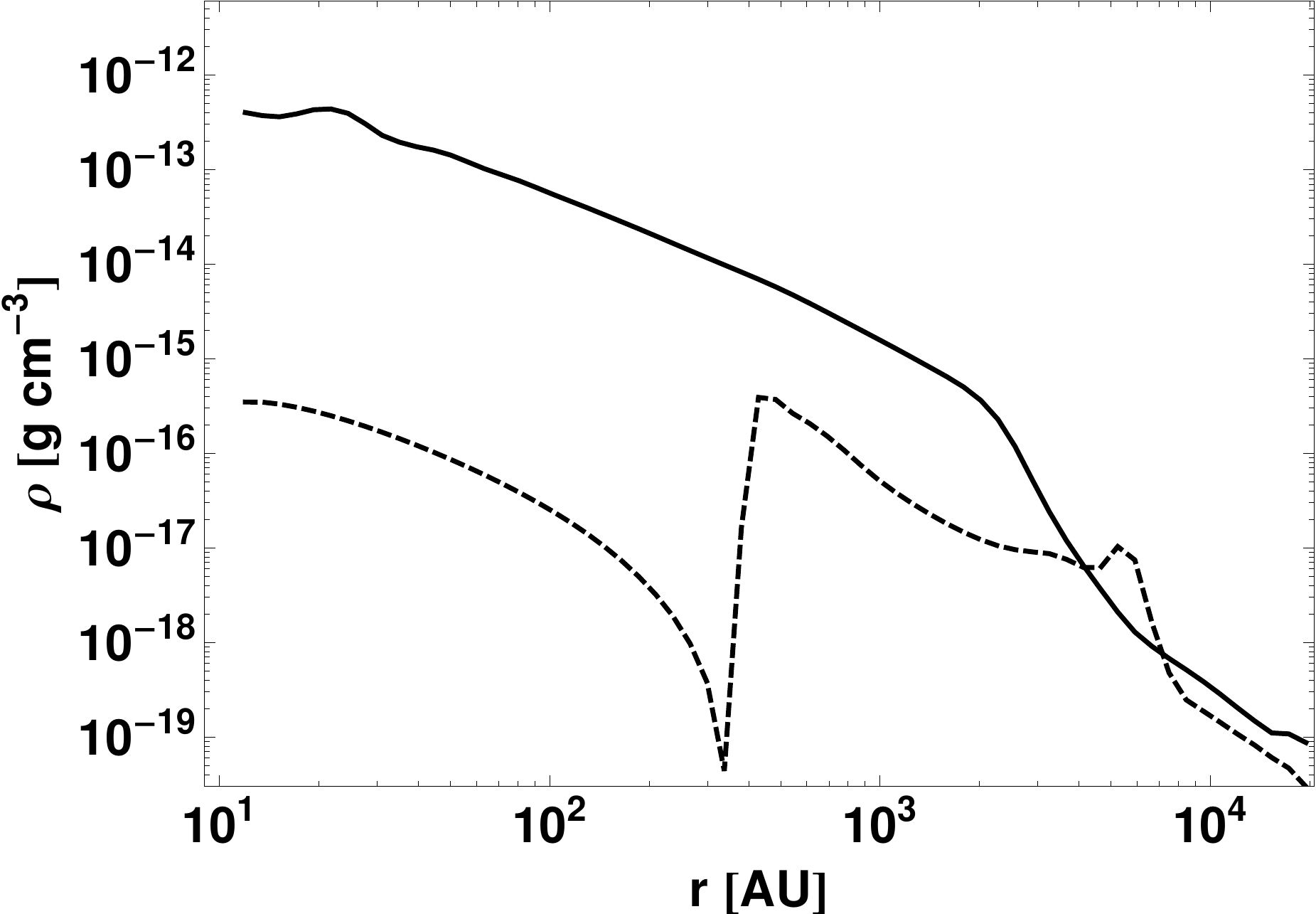}
\label{fig:rho65}
}
\caption{
Radial gas density profiles at the disks' midplanes at different evolution times as indicated.
The solid lines represent the fiducial run; the dashed lines denote the comparison run using a constant gas opacity.
}
\label{fig:rho}
\end{center}
\end{figure*}

\begin{figure*}[p]
\begin{center}
\subfigure[25~kyr]{
\includegraphics[width=0.48\textwidth]{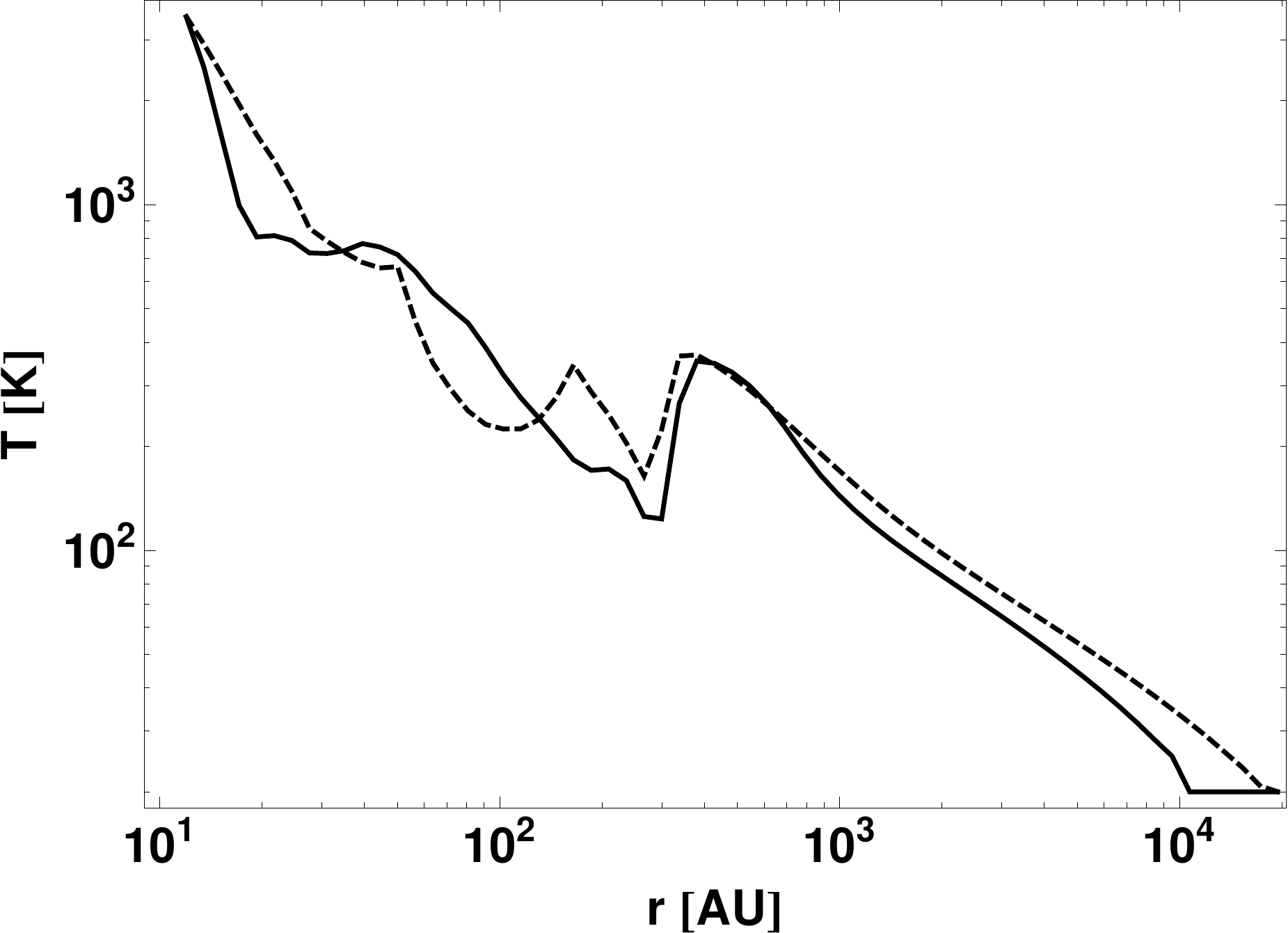}
\label{fig:temp25}
}
\subfigure[40~kyr]{
\includegraphics[width=0.48\textwidth]{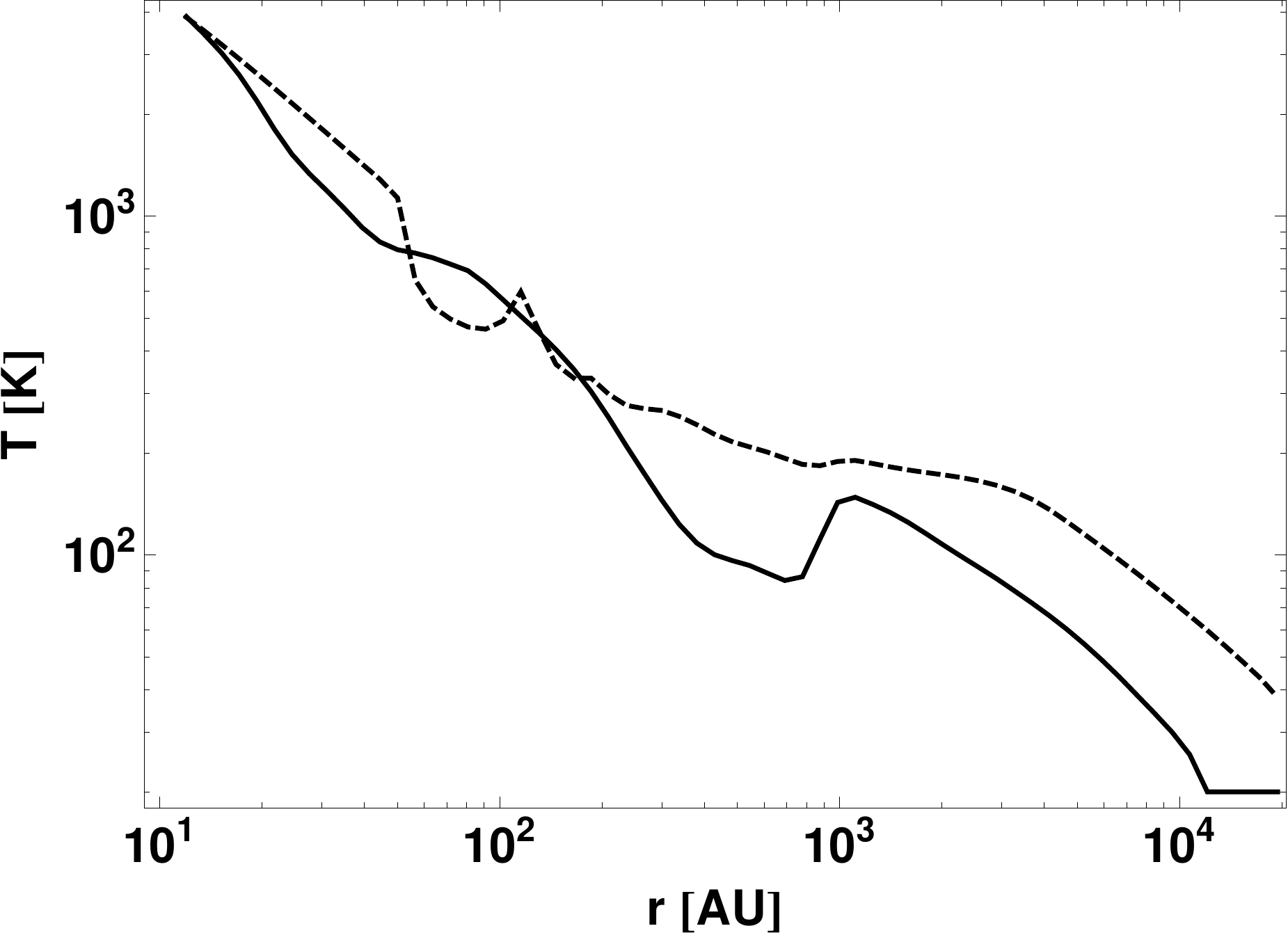}
\label{fig:temp40}
}
\subfigure[50~kyr]{
\includegraphics[width=0.48\textwidth]{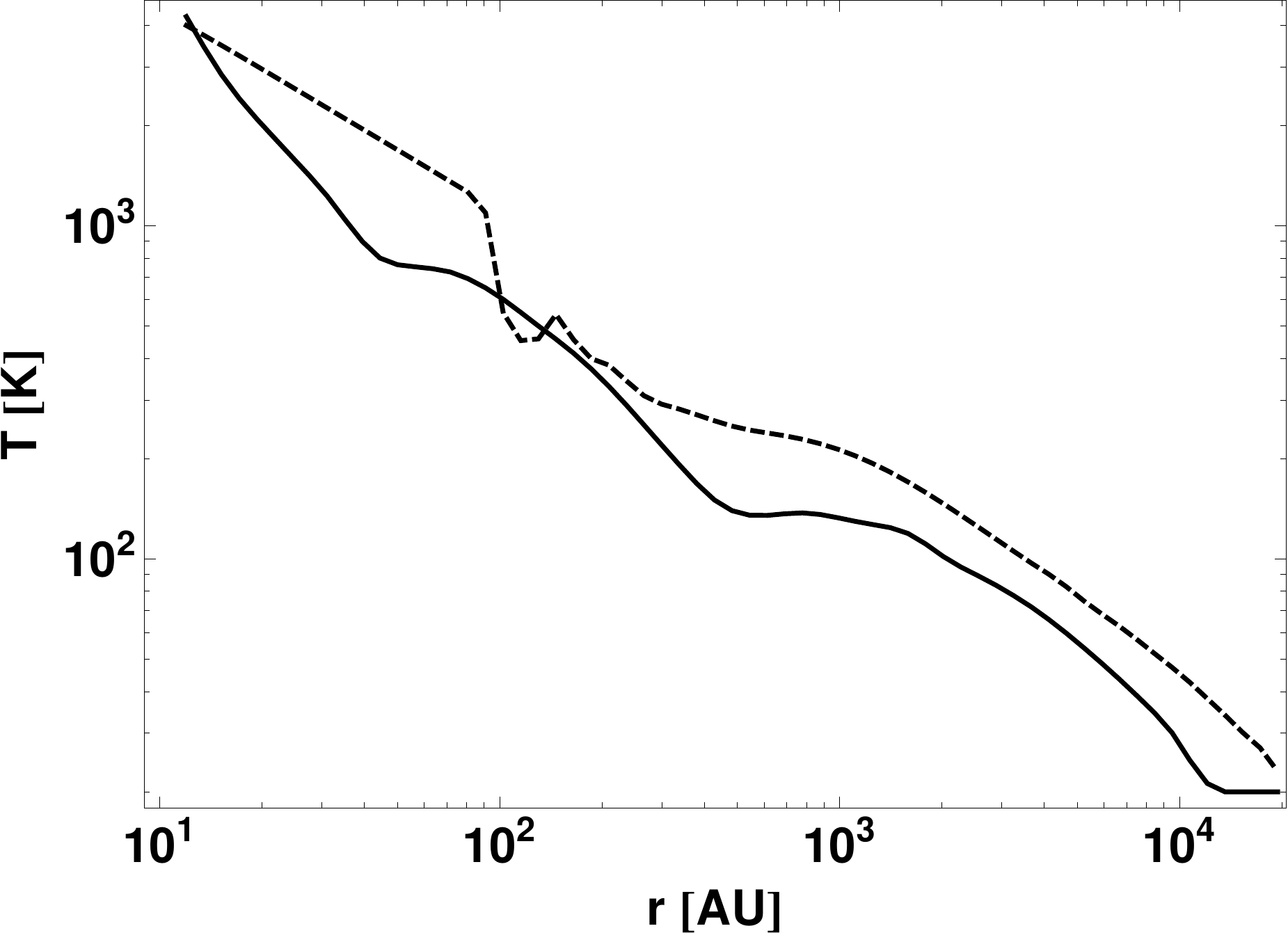}
\label{fig:temp50}
}
\subfigure[55~kyr]{
\includegraphics[width=0.48\textwidth]{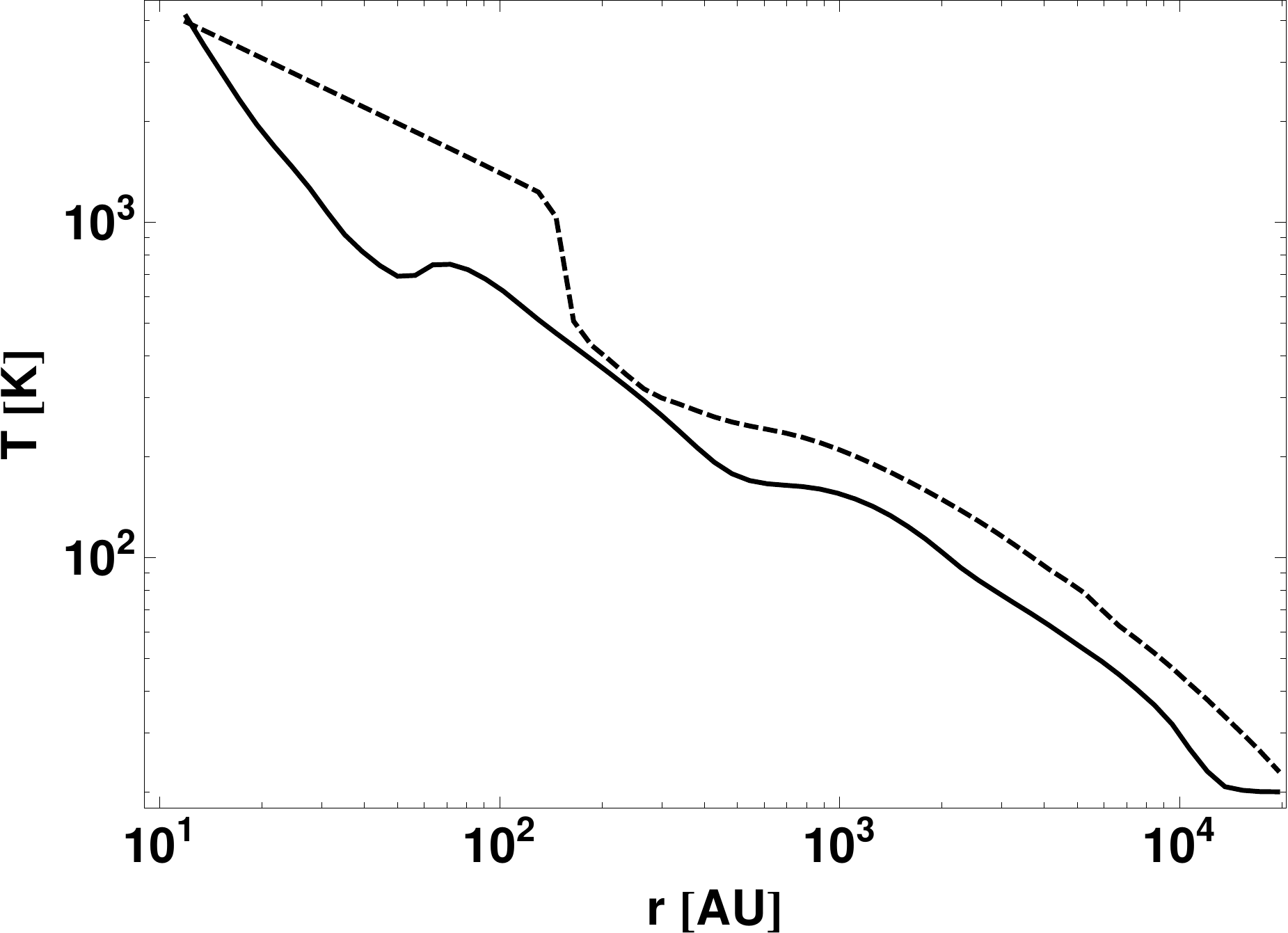}
\label{fig:temp55}
}
\subfigure[60~kyr]{
\includegraphics[width=0.48\textwidth]{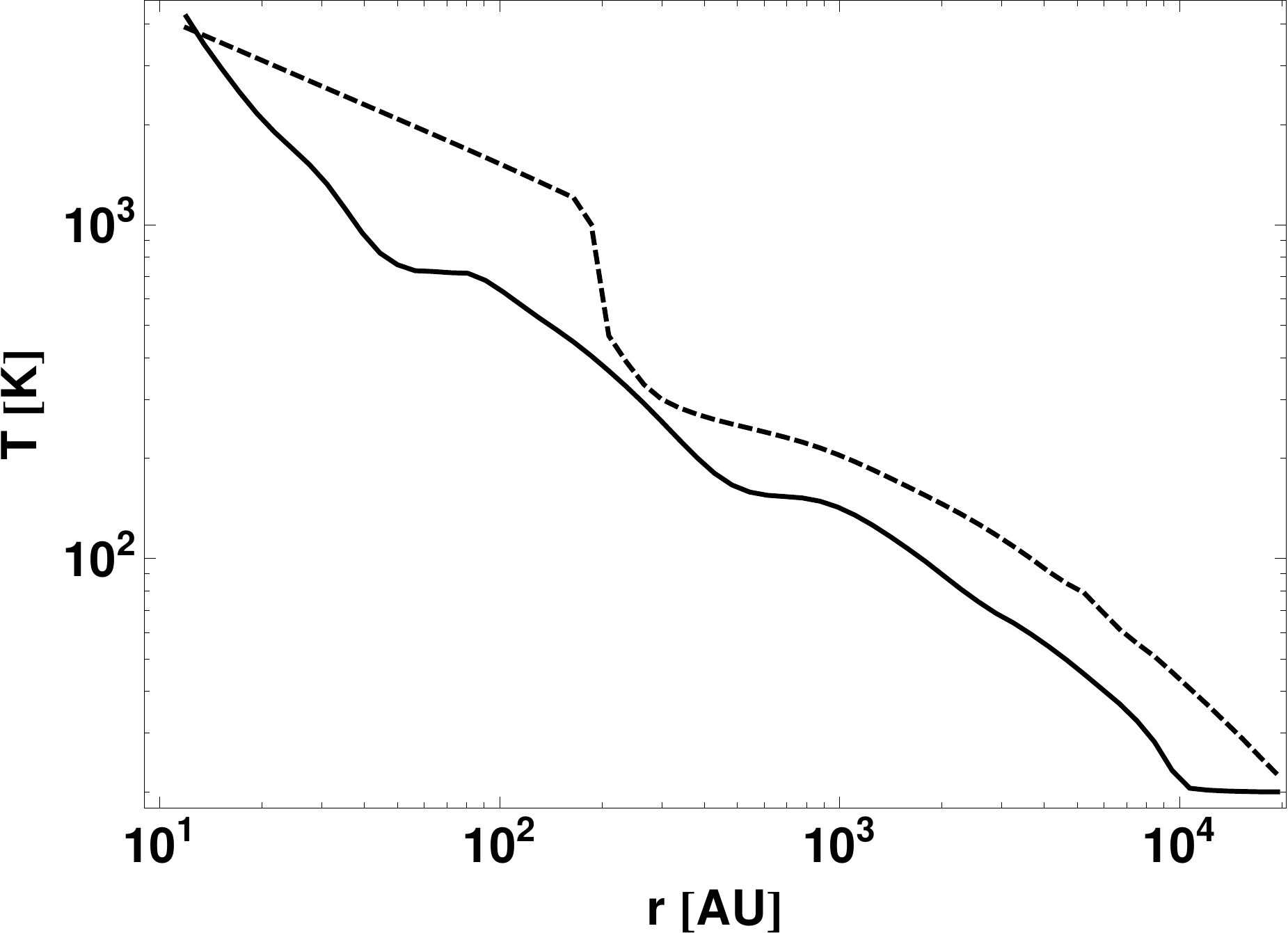}
\label{fig:temp60}
}
\subfigure[65~kyr]{
\includegraphics[width=0.48\textwidth]{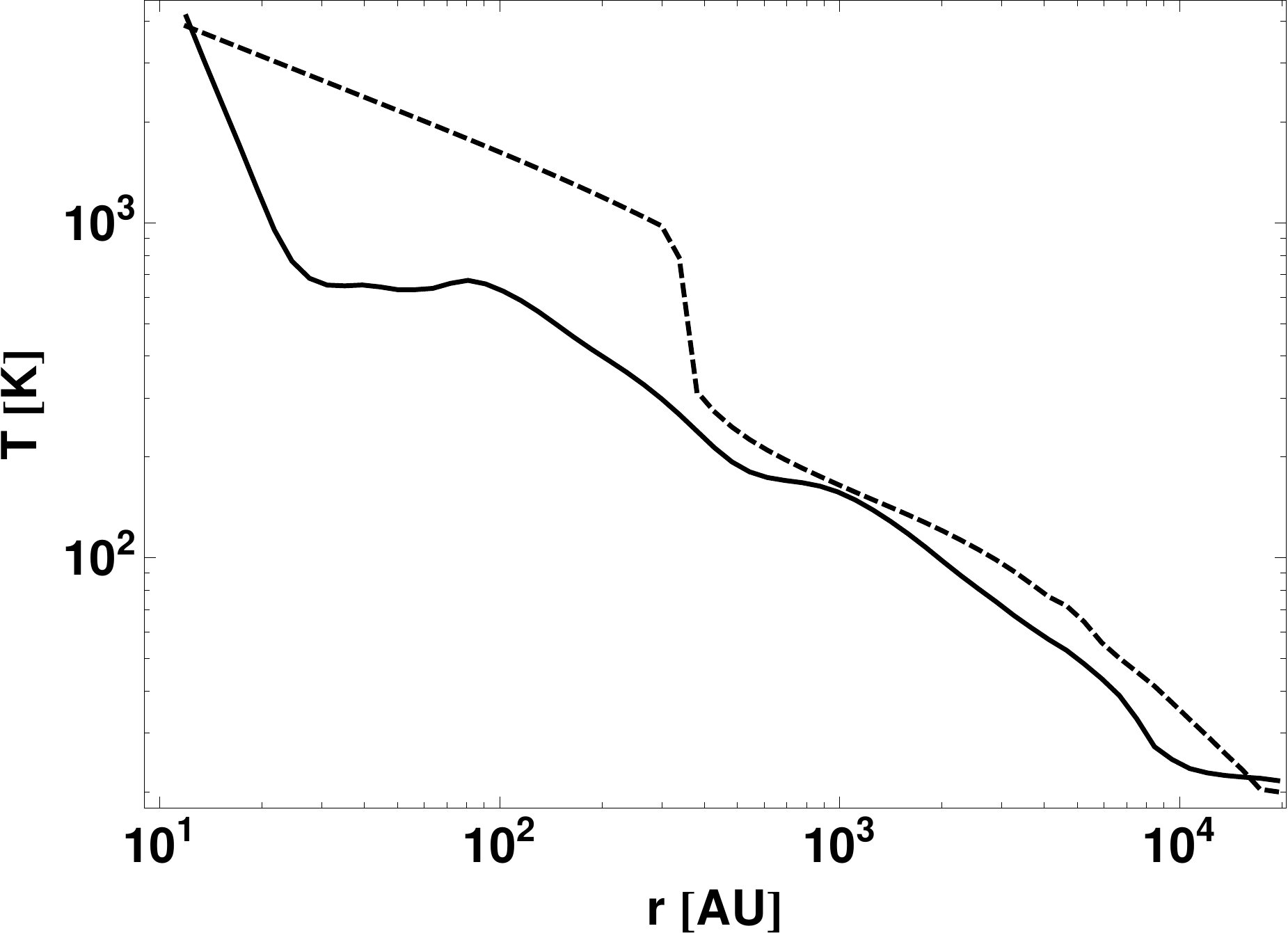}
\label{fig:temp65}
}
\caption{
Radial gas temperature profiles at the disks' midplanes at different evolution times as indicated.
The solid lines represent the fiducial run; the dashed lines denote the comparison run using a constant gas opacity.
}
\label{fig:temp}
\end{center}
\end{figure*}

Disk accretion occurs over a longer period of time in the fiducial run in contrast to the comparison run with lower (constant) gas opacities.  The reason for this is two-fold:
The so-called ``flashlight effect'' is greatly enhanced by the added opacity contribution of the inner dust-free gas with its corresponding shift of dust sublimation to smaller radii in the shielded disk regions.
This prevents the disk from being torn apart by radiative forces for more than $t>200$~kyr, i.e.~most probably for its entire lifetime.
In addition, the accretion disk itself continues to accrete material from the surrounding molecular core for a longer period of time,
because the shielding reduces the ejection of molecular gas by radiative forces.
Because of the reduced accretion onto the disk in the comparison run, it is less massive than the fiducial accretion disk.
As the density in the disk's midplane drops,
{\vONE
the ratio of radiative to gravitational forces increases.}
Radiative forces eventually dominate and drag the remnant disk from the central star, and accretion stops (see Figs.~\ref{fig:rho} and \ref{fig:temp} for a visualization of the density and temperature evolution in the disks' midplane for both simulation runs, respectively).

{\vONE
\subsection{Accretion variability}
The results in this work do not show episodic accretion analogous to that obtained by \citet{Bell:1994p240}, who studied the FU Ori phenomenon using a temperature dependent opacity in the disk.
However, although the opacities used in this study also depend sensitively on the temperature, the variations are not as strong as in the study by \citet{Bell:1994p240}, who examined the thermal instability occurring at the ionization front.
Moreover, what little variability that we obtained in our fiducial run (see Fig.~\ref{fig:Mdot_vs_t}) was reduced when we increased the resolution at the inner disk edge (see discussion of the "convergence run" in the appendix).
We therefore conclude that our accretion variability is numerical rather than of physical origin. 
}

{\vONE
\section{Observational implications}
\subsection{The upper mass limit of stars}
\citet{Kuiper:2010p541} address the question how massive stars can form by disk accretion up to the observed stellar mass range in spite of strong radiation pressure feedback.
In these previous simulations the optically thick dusty disk leads to an anisotropy in the thermal radiation field (the "flashlight effect") enabling steady disk accretion along the radial direction while the majority of thermal radiation escapes in the direction perpendicular to the disk's midplane.
In this study we additionally take into account the dust-free optically thick gaseous disks around massive stars interior to the dusty disks.
This leads to an amplification of the flashlight effect.
For the collapse initial conditions studied here ($M_\mathrm{core} = 100~\mbox{M}_\odot$), the final mass of the star in the fiducial run (taking into account the gas opacities) is 20\% higher than in the comparison run (using a constant gas opacity).
Hence, we conclude that the radiation pressure feedback can efficiently be circumvented by the flashlight effect and
radiation pressure feedback alone does not explain the upper mass limit of stars $M_* \le 150~\mbox{M}_\odot$.

\subsection{Dust sublimation front}
\citet{Kraus:2010p796} report an observation of dust as close as 10~AU around a massive star of about $20~\mbox{M}_\odot$ \citep{Grave:2009p15255}.
Such small dust sublimation radii require efficient shielding of the high-mass stellar irradiation.
As depicted in our simulation analysis in Fig.~\ref{fig:tau}, this shielding can be provided by an optically thick gaseous disk.
The location of the dust sublimation front will furthermore be affected by e.g.~chemical evolution, dust-gas decoupling, magnetic fields, and ionization, which we have not included in this initial study.

\subsection{Disk lifetimes}
Taking into account the absorption by gas in the fiducial run, the feeding of the disk by the large scale envelope allows for disk accretion epochs much longer than computed in the comparison run using a constant gas opacity (at least by a factor of three to four, see Fig.~\ref{fig:Mdot_vs_t}).
We infer from these results that radiation pressure feedback alone seems to be ineffective in limiting the lifetime of accretion disks around massive stars to the observational estimates of $1-2\times10^5~\mbox{yr}$ \citep[see e.g.][]{Cesaroni:2007p8767}.

On the other hand, the lifetime of accretion disks around massive stars might additionally be affected by ionization, photo-evaporation, and fragmentation.
Taking into account the optically thick gas around massive stars yields cooler accretion disks, see Fig.~\ref{fig:temp}.
A cooler disk is more prone to gravitational instabilities and more likely to fragment.
However, fragmentation is not studied in the axially symmetric simulations performed here.

\subsection{Outflow morphologies}
Although the majority of outflows observed are likely to be dominantly driven by magnetic forces, radiative acceleration may play a role for higher luminosity stars in shaping the outflow morphology.
In this study it is shown that the opening angle of the outflow depends on the spatial scales at which it is observed and its point of evolution in time.
In general, the opening angles in the fiducial run are smaller than in the comparison run and are more consistent with observations, whereby molecular outflows tend to have opening angles of $90\degr$ within 50 AU of the star
\citep{Arce:2007p10514}.
}

\section{Limitations \& Outlook}
\label{sect:limitations}
In the low-density bipolar regions the choice of gas opacity is irrelevant because they are extremely optically thin.
In the dust-free parts of the circumstellar disk, however, the correct treatment of gas opacities is important; it 
influences both the large scale shielding of the outer regions of the molecular core and the flashlight effect.  The former
allows accretion onto the circumstellar disk and the latter allows radial flow within the disk.

This study emphasizes the importance of the innermost disk morphology for the radiative feedback efficiency.
We consider it to be a first step in a broader investigation, because many details of the gaseous environment and disk
morphology close to accreting (proto)stars are poorly known.  Other effects, such as ionization, fragmentation and complex magnetic fields, will also influence the local environment.
In fact, a more comprehensive scan of the broad parameter space -- effecting especially the disk morphology such as the initial angular momentum distribution -- will be necessary to further enlighten the quantitative importance of the contribution of the gas phase to the properties of high-mass star forming regions.

Admittedly the enhancement of the flashlight effect detected in this study may be even higher, if the inner sink cell had been smaller
and the spatial resolution in the inner regions had been greater, 
i.e.~if more of the inner gas disk had been included in the computational domain and had been able to contribute to the optical depth of the irradiated accretion disk.
Based on the results of the convergence run with a smaller sink cell radius of $R_\mathrm{min}=5$~AU (see appendix), however,
we are confident that this additional enhancement would not significantly alter our conclusions.

Recent progress of understanding the flashlight effect has been achieved by considering more sophisticated algorithms for
radiation transport in the pre-stellar core collapse simulations.
In particular, the extension toward frequency-dependent stellar irradiation using a ray-tracing method revealed important 
effects such as the stability of radiation-pressure-dominated outflow cavities \citep{Kuiper:2012p1151} and the
validation of the flashlight effect \citep{Kuiper:2010p541}.
Having this in mind, we could expect to achieve a more detailed insight into the radiative feedback of massive stars 
by extending this study to a frequency-dependent treatment of gas opacities.
Unfortunately, however, because of the contributions of molecular line absorption and the complex time-dependent molecular 
chemistry involved, this is not straight-forward.

\acknowledgments
This research project was financially supported by the German Academy of Science Leopoldina within the Leopoldina Fellowship programme, grant no.~LPDS 2011-5.
Author R.~K.~thanks Dmitry Semenov and Mykola Malygin from the Max Planck Institute for Astronomy in Heidelberg for fruitful discussions on theory and numerical computation of frequency-dependent and frequency-averaged gas opacities.
Our work has also profited from critical discussions with Neal Turner and Takashi Hosokawa.
This work was conducted at the Jet Propulsion Laboratory, California Institute of Technology, operating under a contract with the National Aeronautics and Space Administration (NASA).

\bibliographystyle{apj}
\bibliography{Papers.bib}

\appendix
\section{Convergence run}
\begin{figure*}[htbp]
\begin{center}
\includegraphics[width=0.38\textwidth]{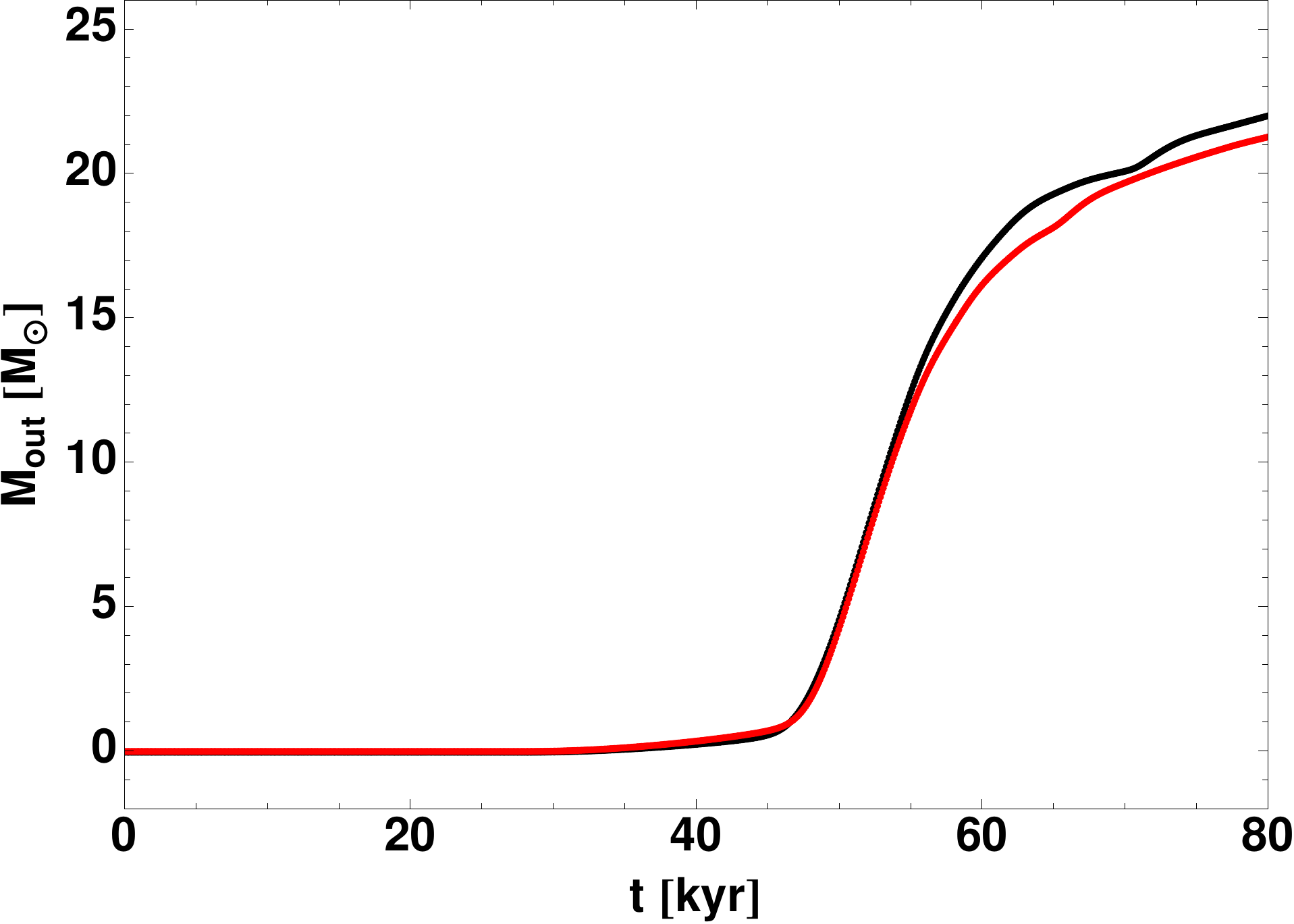}\\
\includegraphics[width=0.38\textwidth]{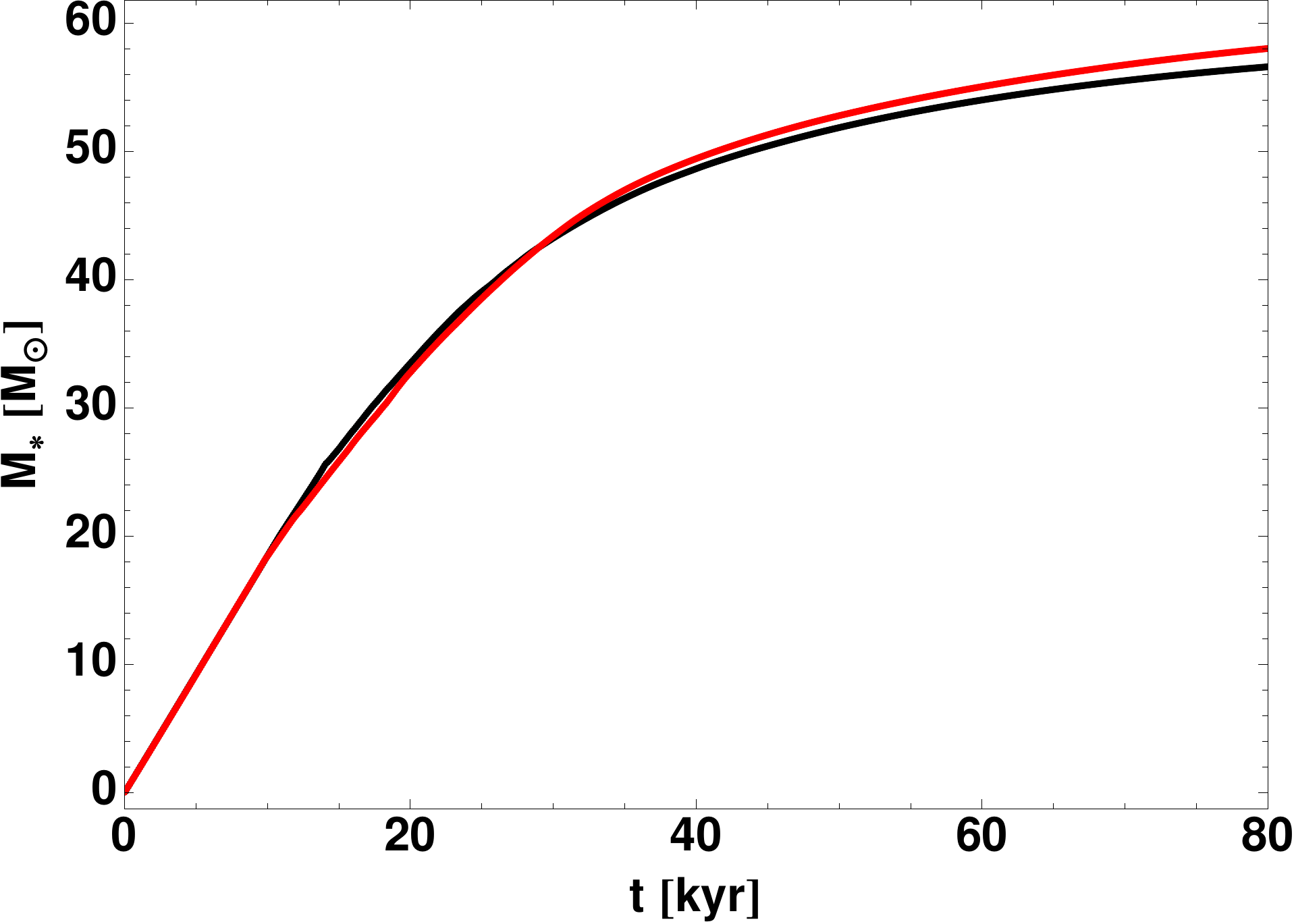}\\
\includegraphics[width=0.38\textwidth]{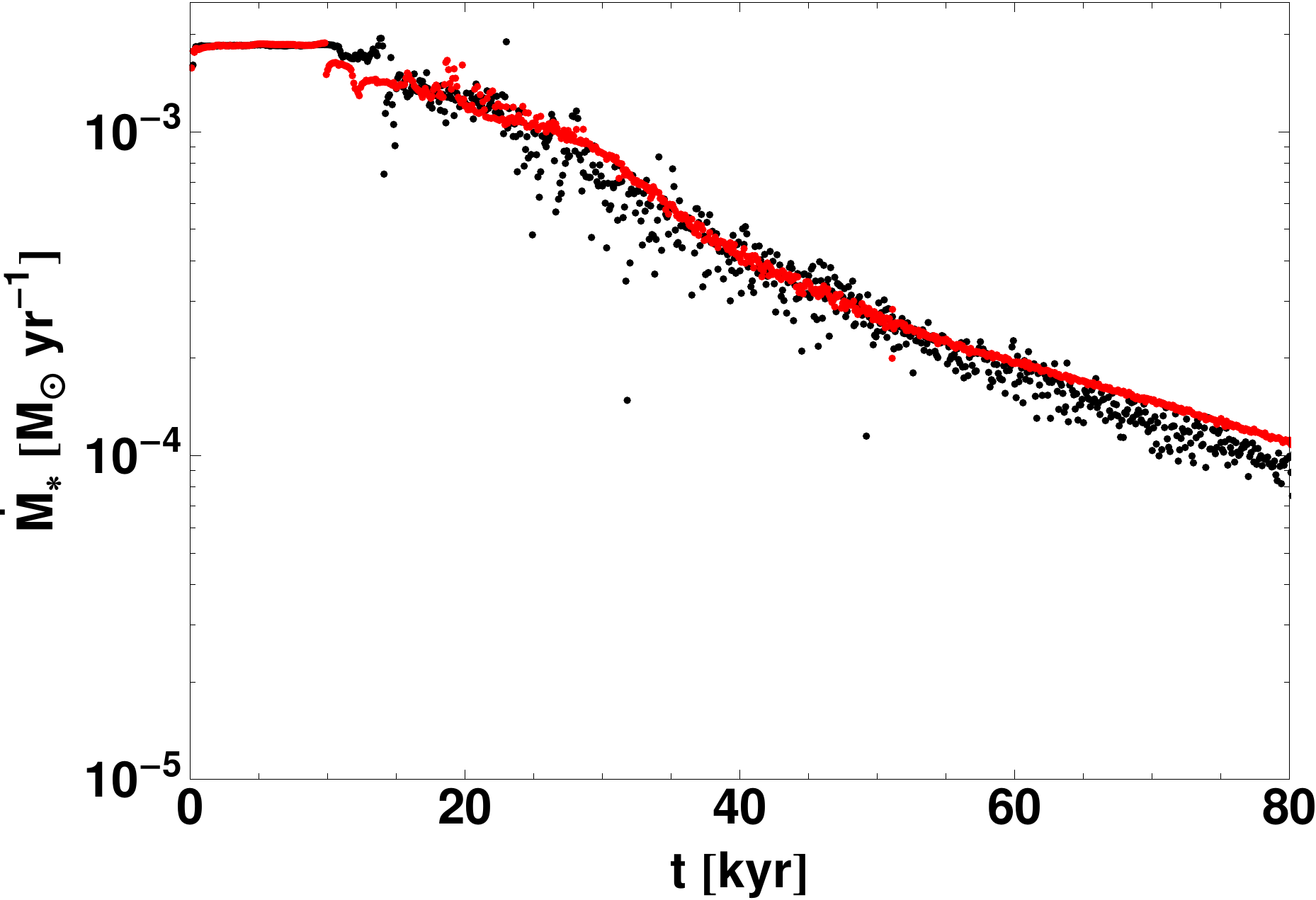}
\caption{
Results of the convergence study using a smaller sink cell radius, which allows a larger part of the innermost gas disk to be included within the computational domain.
Here, 
the mass loss of the pre-stellar molecular core by radiative forces (upper panel),
the mass growth of the central massive star (middle panel), and
the accretion rate onto the central star (lower panel)
as a function of time are given.
The results of the fiducial run using a sink cell size of $R_\mathrm{min}=10$~AU are shown in black.
The results of the convergence run using a sink cell size of $R_\mathrm{min}=5$~AU are shown in red.
}
\label{fig:convergence}
\end{center}
\end{figure*}

To study the dependence of the revealed effects on the size of the sink cell at the inner computational boundary, we repeated the fiducial collapse simulation using a sink cell radius of only $R_\mathrm{min}=5$~AU.
The gas in the region between $r=5$~and~10~AU contributes to the total optical depth of the inner gas disk and therefore could alter the results obtained in this study.
Fig.~\ref{fig:convergence} compares the most relevant physical properties of the collapsing system to the fiducial run.
Indeed, as expected, the increase of the total optical depth to stellar illumination further enhances the flashlight effect and the shielding of the large scale mass reservoir in the molecular core.
In addition, the accretion rate into the inner sink cell shows less (numerical) variability due to the improved polar resolution at the inner computational boundary.

\end{document}